\documentclass[10pt]{article}
\pdfoutput=1
\usepackage{graphicx}
\usepackage{hyperref}
\usepackage{amssymb, amsfonts, amsmath}
\usepackage{cite}
\usepackage{subfigure}
\usepackage{authblk}
\usepackage{color}

\newcommand{\g}[1]{\gamma_{#1}} % gamma matrices (covariant index)
\renewcommand{\l}{\left}
\renewcommand{\r}{\right}
\newcommand{\bra}[1]{\left< #1 \right|} % defines the Bra
\newcommand{\ket}[1]{\left| #1 \right>} % defines the Ket

\newcommand{\gev}{\,\mathrm{GeV}}

\newcommand{\mev}{\,\mathrm{MeV}}
\newcommand{\fm}{\,\mathrm{fm}}

\newcommand{\order}[1]{\mathcal{O}\l({#1}\r)}
\newcommand{\SU}[1]{\mathrm{SU}\l(#1\r)}

\newcommand{\stat}{\mathrm{stat}}
\newcommand{\sys}{\mathrm{sys}}

\newcommand{\iso}{\mathrm{isov}}

\newcommand{\Ftilde}{\tilde{F}}
\newcommand{\ftilde}{\tilde{f}}
\newcommand{\xvec}{\vec{x}}
\newcommand{\epow}[1]{\mathrm{e}^{#1}}
\newcommand{\Nspatial}{N_{s}}

\title{\textbf{Position space method for the nucleon magnetic moment in lattice QCD} \\[1cm]
\begin{center}
 \includegraphics[draft=false,width=0.13\linewidth]{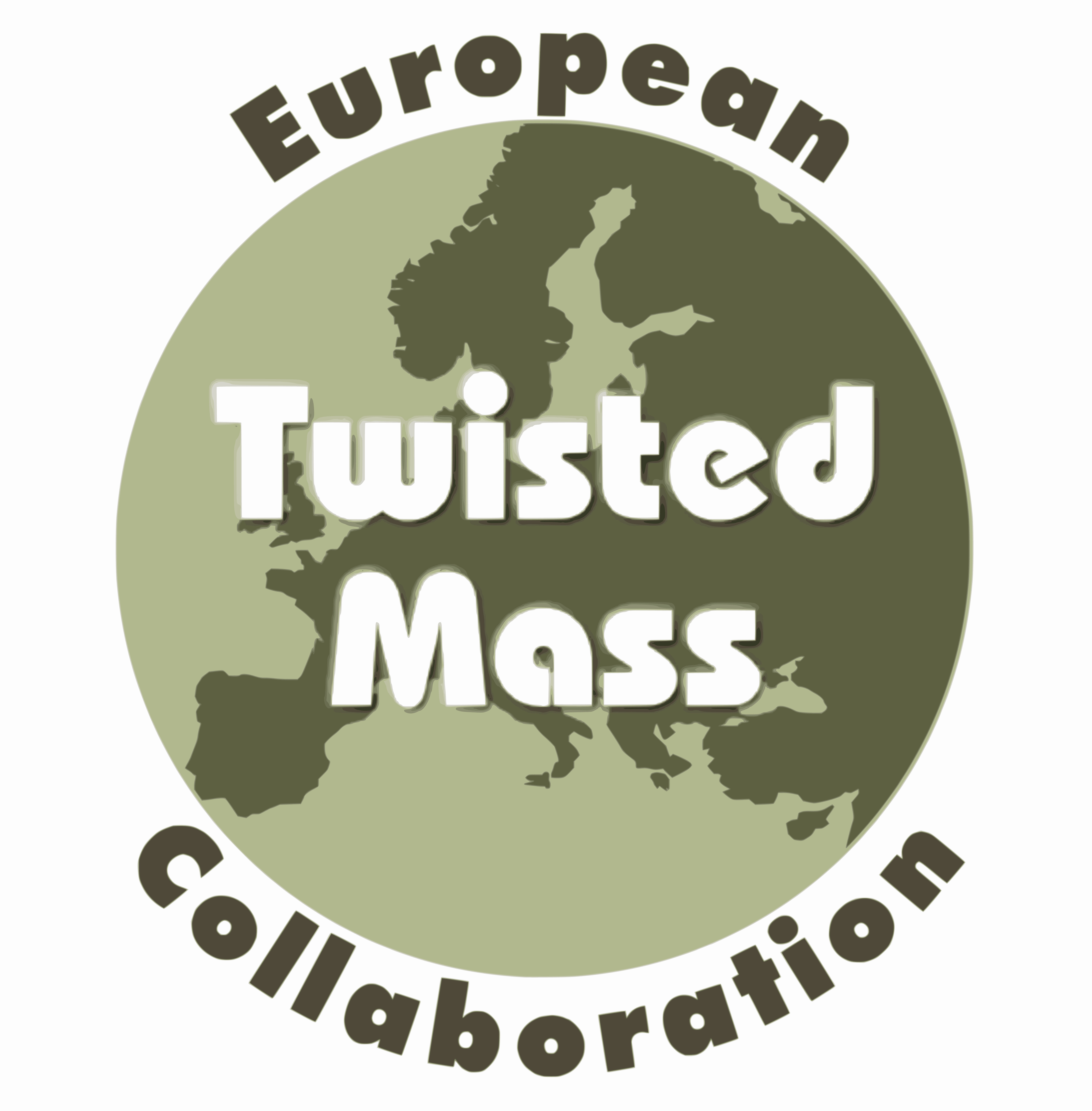}
 \end{center}
}

\author[(a,b)]{Constantia~Alexandrou}
\author[(a,b)]{Martha~Constantinou}
\author[(b)]{Giannis~Koutsou}
\author[(a,c)]{Konstantin~Ottnad}
\author[(b,c)]{Marcus~Petschlies}

\affil[(a)]{Department of Physics, University of Cyprus, P.O. Box 20537, 1678 Nicosia, Cyprus}
\affil[(b)]{Computation-based Science and Technology Research Center, 20 C. Kavafi Street, 2121 Nicosia, Cyprus}
\affil[(c)]{Institut f\"ur Strahlen- und Kernphysik (Theorie), Nussallee 14-16 and Bethe Center for Theoretical Physics, Nussallee 12, Universit\"at Bonn, D-53115 Bonn, Germany}

\begin{document}
 \maketitle
 \begin{abstract}
  \noindent The extraction of the magnetic form factor of the nucleon at zero momentum transfer is usually performed by adopting a parametrization for its momentum dependence and fitting the results obtained at finite momenta. We present a position space method that allows us to remove the momentum prefactor in the form factor decomposition and hence compute the magnetic form factor directly at zero momentum without the need to assume a functional form for its momentum dependence. The method is explored on one ensemble using $N_f=2+1+1$ Wilson twisted mass fermions with a light quark mass corresponding to $M_\pi=373$~MeV and a lattice spacing of $a\approx0.082\fm$. We obtain results for the isovector magnetic moment and for the proton and neutron magnetic moments. The value we find for the isovector magnetic moment is larger as compared to fitting the form factor at the discrete values of the lattice momentum transfers using a dipole Ansatz, bringing it closer to the experimental value.
 \end{abstract}

\clearpage
\section{Introduction}
The proton electromagnetic form factors have been studied extensively for many years as probes of the proton structure. Polarization experiments can measure directly the ratio of the proton electric to magnetic form factor, $\mu_p G_E^p /G^p_M$ and revealed a qualitatively different behavior than the traditional Rosenbluth separation. Future experiments are being planned to study both the small and the large momentum dependence of these form factors. Lattice QCD provides a well-suited framework for the determination of these quantities. This is particularly true nowadays with the availability of simulations with quark masses fixed to their physical values. In the lattice QCD formulation the form factors are evaluated at discrete values of the momentum transfer. Access to low momentum transfers is crucial and will enable a better determination of the magnetic moment of the proton as well as the root mean squared (r.m.s.) radius of the transverse electric and magnetic quark distribution in the nucleon. Computing the electric r.m.s. radius of the proton is particularly important due to the existing discrepancy between the muonic determination yielding a value of $r_p = 0.84184(67)$~fm~\cite{Antognini:1900ns} and the electron scattering and hydrogen spectroscopy value of $r_p=0.8751(61)$~fm~\cite{Mohr:2015ccw}. To access small momentum transfers we need a lattice with large spatial extent, $L$, since the smallest available momentum in a finite box is $2\pi/L$ for periodic boundary conditions. In order to extract the magnetic moment one needs to determine the magnetic form factor at zero momentum transfer. Because the kinematical factor multiplying the magnetic form factor vanishes at zero momentum transfer one cannot extract it directly at zero momentum. Instead one resorts to an extrapolation using an ansatz to fit its momentum dependence. Similarly, to determine the radii one needs the derivative of the form factors in the limit of zero momentum transfer and thus an ansatz for the momentum dependence. Given that on a finite lattice only discrete values of the momentum are allowed, such extrapolations can introduce uncontrolled errors. \par

In this article we investigate an approach that allows us to remove momentum prefactors in the form factor decomposition of a matrix element computed within lattice QCD. Methods of this kind have been applied already in various forms in the study of the hadronic vacuum polarization on the lattice \cite{Allison:2008xk,Bernecker:2011gh,Feng:2013xsa,Francis:2013qna,Malak:2015sla}. However, in contrast to the inclusive process described by hadronic vacuum polarization, the extraction of a form factor associated with the ground state of a correlation function requires modifications and further development of these methods. \par

We first review a variant of the method applied some time ago for the elimination of momentum prefactors~\cite{Wilcox:2002zt}. We discuss the improvements made to avoid the residual time dependence that did not allow the application of  the method to lattice three point functions. Our new method was first considered in Ref.~\cite{Alexandrou:2014exa} and more recently applied in an exploratory study of the neutron electric dipole moment~\cite{Alexandrou:2015spa,Alexandrou:2015ttm}. We describe here the technical details of the method and discuss the application to the nucleon magnetic moment, for which it was originally developed. The method can also be extended to the case of radii for which results will be reported in the future. \par

Before we apply our approach to the magnetic form factor we first  demonstrate its validity  by applying it to evaluate the isovector electric Sachs form factor at zero momentum transfer. The latter is by definition equal to one if the lattice conserved electromagnetic current is used and, in addition, it can be extracted directly from the lattice matrix element without the need to deal with a momentum factor in the decomposition. It thus provides a very suitable test case. After demonstrating the suitability of our method with the electric form factor we apply it to compute the magnetic form factor at zero momentum transfer, which otherwise requires a fit ansatz to extrapolate the lattice data to zero momentum. We find that the value extracted is larger than what a dipole fit to the form factor would yield. We finally discuss our conclusion and future applications of our method. \par

\section{Electromagnetic matrix element of the nucleon}
In this study we consider the electromagnetic matrix element of the nucleon
\begin{equation}
 \bra{N(p_f,s_f)} J_\mu \ket{N(p_i,s_i)} = \frac{m_N}{\sqrt{E_N\l(\vec{p}_f\r)E_N\l(\vec{p}_i\r)}} \bar{u}(p_f,s_f) \l[ \gamma_\mu F_1(q^2) + \frac{i\sigma_{\mu\nu} q_\nu}{2m_N} F_2(q^2) \r] u(p_i,s_i) \,,
 \label{eq:matrix_element}
\end{equation}
where the momentum transfer squared in Minkowski space is given by $q^2=\l(p_f-p_i\r)^2$ while $p_i$, $s_i$ and $p_f$, $s_f$ denote momentum and spin of initial and final state, respectively. In the following the final state is always assumed to be produced at rest, hence initial and final momenta fulfill $\vec{q} = \vec{p}_f - \vec{p}_i = -\vec{p}_i$. Moreover, assuming exact $\SU{2}$ isospin symmetry the local electromagnetic current
\begin{equation}
 J^\mathrm{em}_\mu = \frac{2}{3} \bar{u} \g{\mu} u - \frac{1}{3} \bar{d} \g{\mu} d \,,
 \label{eq:elmag_current}
\end{equation}
satisfies the following relation 
\begin{equation}
 \bra{p} J^\mathrm{em}_\mu \ket{p} - \bra{n} J^\mathrm{em}_\mu \ket{n} = \bra{p} \bar{u} \g{\mu} u - \bar{d} \g{\mu} d \ket{p} \equiv \bra{p} J^\iso_\mu \ket{p} \,,
 \label{eq:isospin_current}
\end{equation}
where the isovector electromagnetic current $J^\iso_\mu$ has been introduced, and $\ket{p}$, $\ket{n}$ refer to proton and neutron states, respectively. Using the isovector combination instead of the electromagnetic current allows us to avoid quark disconnected diagrams, which would otherwise contribute to the matrix element in Eq.~(\ref{eq:matrix_element}). Furthermore, we replace the local current in our lattice computations by the corresponding Noether current, eliminating the need for additional renormalization. \par

In Euclidean spacetime the momentum transfer squared is given by $Q^2=-q^2$ and the corresponding Dirac and Pauli form factors $F_1\l(Q^2\r)$ and $F_2\l(Q^2\r)$ are related to the electric and magnetic Sachs form factors by
\begin{align}
 G_E\l(Q^2\r) &= F_1\l(Q^2\r) - \frac{Q^2}{4m_N^2} F_2\l(Q^2\r) \,, \notag \\
 G_M\l(Q^2\r) &= F_1\l(Q^2\r) + F_2\l(Q^2\r) \,.
\end{align}
On the lattice, we consider spin-projected two-point and three-point functions
\begin{align}
 C^{2pt}(t,\vec{q}) &= \Gamma^{\alpha\beta}_0 \langle  J^\alpha_N(\vec{q},t_f) {\overline J}_N^\beta(\vec{q},t_i) \rangle  \,, \\
 C^{3pt}_\mu(t,\vec{q},\Gamma_\nu) &= \Gamma^{\alpha\beta}_\nu \langle J_N^\alpha(\vec{0},t_f) J^{\rm em}_\mu(\vec q, t) {\overline J}_N^\beta({\vec p}_i,t_i)\rangle \,,
 \label{eq:3pt_function_interpolator}
\end{align}
where $J_{N,\alpha}$ denotes a suitable interpolating field for the desired nucleon state. The projectors relevant for our purposes are defined by $\Gamma_0 = \frac{1}{2} \l(1+\g{0}\r)$ and $\Gamma_k = \frac{1}{4} \Gamma_0 i \g{5}\g{k}$. Unknown overlap factors due to the spin projection are canceled in the optimized ratio
\begin{equation}
 R_\mu(t_f,t,\vec{q},\Gamma_\nu) = \frac{C_\mu^{3pt}(t_f,t,\vec{q},\Gamma_\nu)}{C^{2pt}(t_f,\vec{0})} \sqrt{\frac{C^{2pt}(t_f-t,\vec{q}) C^{2pt}(t, \vec{0}) C^{2pt}(t_f, \vec{0})}{C^{2pt}(t_f-t, \vec{0}) C^{2pt}(t, \vec{q}) C^{2pt}(t_f, \vec{q})}} \,,
 \label{eq:ratio}
\end{equation}
where $t_f$, $t$ denote the (fixed) sink and (running) insertion time slices, respectively. Without loss of generality, the source timeslice $t_i$ has been set to zero and dropped entirely. The required three-point functions are computed using sequential inversions through the sink \cite{Dolgov:2002zm} which gives access to the full $Q^2$--dependence. At large Euclidean times $t$ and $t_f-t$, the ground state is expected to dominate the ratio and $R_\mu(t_f,t,\vec{q},\Gamma_\nu)$ approaches a plateau, i.e.
\begin{equation}
 \lim\limits_{t\rightarrow\infty} \ \lim\limits_{t_f-t\rightarrow\infty} R_\mu(t_f,t,\vec{q},\Gamma_\nu) = \Pi_\mu\l(\vec{q}, \Gamma_\nu\r) \,.
 \label{eq:plateau}
\end{equation}
Fitting a constant to this plateau finally allows us to extract the isovector Sachs form factors by employing an appropriate choice of projectors and insertion indices
\begin{align}
 \Pi_0\l(\vec{q},\Gamma_0\r) &= -C\frac{E_N\l(\vec{q}\r)+m_N}{2m_N} G_E\l(Q^2\r) \label{eq:G_E} \,, \\
 \Pi_i\l(\vec{q},\Gamma_0\r) &= -C\frac{i}{2m_N} q_i G_E\l(Q^2\r) \label{eq:q_G_E} \,, \\
 \Pi_i\l(\vec{q},\Gamma_k\r) &= -C\frac{1}{4m_N} \epsilon_{ijk} q_j G_M\l(Q^2\r) \label{eq:q_G_M} \,,
\end{align}
where we have introduced the kinematic factor $C=\sqrt{\frac{2m_N^2}{E_N\l(\vec{q}\r)\l(E_N\l(\vec{q}\r)+m_N\r)}}$.  From Eq.~(\ref{eq:G_E}) it is obvious that the isovector electric moment $G_E\l(0\r)=1$ can be extracted directly from a lattice calculation, whereas for the case of the anomalous magnetic moment $G_M\l(0\r)$ no relation without a multiplicative momentum factor exists. A standard method to obtain an estimate for $G_M\l(0\r)$ is to assume a fit ansatz to describe the data at nonzero momentum transfer and use the fitted parameters to extrapolate to zero momentum. A simple and common choice is a dipole fit of the form
\begin{equation}
 G_M(Q^2) = \frac{G_M\l(0\r)}{\l(1+\frac{Q^2}{m_{G_M}^2}\r)^2}\,,
 \label{eq:dipole_fit}
\end{equation}
where $G_M\l(0\r)$ and the dipole mass $m_{G_M}$ are fit parameters. However, any such approach introduces a model dependence, which given the discrete nature of $Q^2$ in a lattice calculation can be problematic. In fact, many different ans\"atze have been discussed in the literature; for a recent review we refer to \cite{Punjabi:2015bba}. 

Here we follow a different, model-independent approach that relies on the treatment of correlation functions in position space. This will allow us to access $G_M\l(0\r)$ and similar quantities from lattice data without the need for an explicit parametrization for the momentum dependence of the form factor. \par

\section{Position space methods}
Assuming continuous (infinite) Euclidean spacetime it is possible to remove the $q_j$--factor in Eq.~(\ref{eq:q_G_M}) by application of a derivative. Taking the $Q^2\rightarrow0$ limit we find
\begin{equation}
 \lim\limits_{Q^2\rightarrow0} \frac{\partial}{\partial q_j} \Pi_i\l(\vec{q}, \Gamma_k\r) = -\frac{1}{4m_N} \, \epsilon_{ijk} G_M\l(0\r) \,.
 \label{eq:derivative}
\end{equation}
In principle, a corresponding procedure can be defined in a finite Euclidean spacetime where the momenta are discrete. However, there is no unique prescription for this, as the definition of a derivative on the lattice is not unambiguous. Furthermore, additional subtleties may occur when taking the large Euclidean time and infinite volume limits, which are not interchangeable in general. This issue was first discussed in Ref.~\cite{Wilcox:2002zt}, where it has been shown that naively applying a momentum derivative on the lattice does not yield the correct result even in the infinite volume limit, i.e. Eq.~(\ref{eq:derivative}) is not reproduced correctly. In the following subsection we briefly revisit this result and discuss how to avoid its implications when approaching the large Euclidean time and infinite volume limit. In a next step we derive a method that avoids the aforementioned issues altogether.

\subsection{Direct application of the momentum derivative}
\label{subsec:direct_application_of_the_momentum_derivative}
For continuum quantities in infinite volume one can eliminate the momentum factor $q_j$--factor in Eqation~(\ref{eq:q_G_M}) by applying a partial derivative acting on the ratio $R_i$ in Eq.~(\ref{eq:ratio}) and write 
\begin{align}
 \lim\limits_{Q^2\rightarrow0} \frac{\partial}{\partial q_j} R_i(t_f,t,\vec{q},\Gamma_k) &= 
 \frac{1}{C^\mathrm{2pt}(t_f,\vec{0})}\,
 \lim\limits_{Q^2\rightarrow0} \,
 \frac{\partial}{\partial q_j}\,C_i^\mathrm{3pt}(t,\vec{q},\Gamma_k), \nonumber \\
 &= \,\frac{1}{C^\mathrm{2pt}(t_f,\vec{0})} \cdot \int dx_j d^2x_{\bot j} i x_j C_i^\mathrm{3pt}(t, \vec{x}) \,.
 \label{eq:continuum_derivative0}
\end{align} 
where we have introduced the three-point function $C_i^\mathrm{3pt}(t, \vec{x})$ in position space and for fixed $j \in \left\{ 1,2,3 \right\}$ two-dimensional integration is over the two components of $\vec x$ perpendicular to $j$. Any term involving a derivative of a two-point function in the above expression vanishes exactly, hence there is no contribution from the square-root of two-point functions in Eq.~(\ref{eq:ratio}). One can naively write the discretized version of the second line in Eq.~(\ref{eq:continuum_derivative0}) to leading order in the lattice spacing and for infinite spatial lattice length $L$ equate it to the partial derivative of the ratio. One would then write
\begin{equation}
   \lim\limits_{Q^2\rightarrow0} \frac{\partial}{\partial q_j} R_i(t_f,t,\vec{q},\Gamma_k)  =  \lim\limits_{L \to \infty} \,\frac{1}{C^\mathrm{2pt}(t_f,\vec{0})} \cdot a^3 \sum\limits_{x_j=-L/2+a}^{L/2-a} \left( \sum\limits_{\vec x_{\bot j}=0}^{L-a} i x_j  C_i^\mathrm{3pt}(t, \vec{x}) \right) \,,
 \label{eq:continuum_derivative}
\end{equation}
where the sum over $\vec x_{\bot j}$ means the sum over the two components of $\vec x$ perpendicular to $j$ over the complete range $\left\{ 0,\ldots,L-a \right\}$. Note that in Eq.~(\ref{eq:continuum_derivative}) we have interchanged the partial momentum derivative and the limit of zero momentum an operation that is well-defined in infinite volume. \par

In finite volume the above expression approximates the momentum-space derivative of a $\delta$-distribution in infinite volume,
\begin{align}
 a^3 \hspace{-0.4cm}\sum_{x_j = -L/2 + a}^{L/2 - a} \hspace{-0.1cm} \left( \sum_{\substack{\vec x_{\bot j}=0 }}^{L -a} 
 \hspace{-0.1cm} i x_j C_i^\mathrm{3pt} (t_f, t, \vec{x}, \Gamma_k) \right) \hspace{-0.2cm} \ 
 &= \ \hspace{-0.2cm} {\frac{1}{V} \sum_{\vec{k}}  
 \hspace{-0.1cm} \left( a^3 \hspace{-0.4cm}\sum_{x_j = -L/2 + a}^{L/2 - a}  
 \hspace{-0.1cm} 
 \left( \sum_{\substack{\vec x_{\bot j}=0 }}^{L -a} i x_j \exp\left( i \vec{k} \vec{x}  \right)\right) 
    \right) \,
    C_i^\mathrm{3pt} (t_f, t, \vec{k}, \Gamma_k)\,,} \nonumber \\ 
& \stackrel{L\rightarrow \infty}{\longrightarrow}\frac{1}{\left( 2 \pi  \right)^3} \int d^3 \vec{k} \frac{\partial}{\partial k_j} \delta^{(3)}(\vec{k})\, C_i^\mathrm{3pt} (t_f, t, \vec{k}, \Gamma_k) \,,
 \label{eq:delta_approximation}
\end{align}
which implies a residual $t$--dependence $C_i^\mathrm{3pt}(t_f, t,\vec{q},\Gamma_k) \sim \exp(-\Delta E t)$, where $\Delta E = E_N(\vec{q}) -m_N$ is the momentum transfer between final and initial state and $\Delta E\rightarrow 0$ only for $L\rightarrow \infty$. In any numerical study this dependence must either be fitted (if statistical errors are small or the value of $L/a$ not large enough) as it was done in Ref.~\cite{Alexandrou:2014exa}, or at least taken into account as an additional systematic uncertainty~\cite{Alexandrou:2015spa}. \par 

From Eqs. (\ref{eq:continuum_derivative}) and (\ref{eq:delta_approximation}) we thus formally define our lattice estimator in position space for the form factor at zero momentum as $\mathcal{F}(t,t_f,L)$ given by
\begin{equation}
  \mathcal{F}(t,t_f,L)
  = \frac{2 m_N}{3} \cdot \frac{1}{C^{\mathrm{2pt}}( t_f,\vec 0 )}\,\sum\limits_{i,j,k=1}^{3}\,
  \epsilon_{jik}\, a^3 \sum\limits_{x_j = -L/2+a}^{L/2-a}\,\sum\limits_{x_i,x_k=0}^{L-a}\,
  x_j\,C^{\mathrm{3pt}}_i\left( t_f,t,\xvec,\Gamma_k \right)
  \label{eq:G_M_position_space_sum0}
\end{equation}
which taking  the usual limits of infinite lattice size, source-sink time separation and time separation of the current insertion from source and sink yields the form factor at $Q^2=0$:
\begin{equation}
  G_M\left( 0 \right) = \lim\limits_{t\rightarrow\infty} \, \lim\limits_{t_f-t\rightarrow\infty}\, \lim\limits_{L \to\infty}\,
  \mathcal{F}(t,t_f,L) \,.
 \label{eq:G_M_position_space_sum}
\end{equation}
The  time and volume dependence of $\mathcal{F}$ in Eq.~(\ref{eq:G_M_position_space_sum}), which we calculate on the lattice, play a crucial role in the direct application of the momentum derivative. \par

To facilitate the discussion, for fixed value of $j$ and $x = x_j$ we define 
\begin{equation}
  F\left(t, x \right) \equiv \frac{1}{C^{\mathrm{2pt}} ( t_f, \vec 0 )}  \,
    \epsilon_{ijk} \,\sum\limits_{x_i,x_k=0}^{L-a}\,C^{\mathrm{3pt}}_i\left( t_f,t,\xvec,\Gamma_k  \right)
  \label{eq:F_n}
\end{equation}
and its Fourier transform $\Ftilde$ in momentum space. We assume a sufficiently large source-sink time separation $t_\mathrm{sep}=t_f - t_i$, such that the desired nucleon states dominate. Both $F$ and $\Ftilde$ will depend on the insertion time $t$, though this is not explicitly denoted in each step.

Using trigonometric interpolation, we can evaluate $\Ftilde$ for intermediate momentum $k$ via
\begin{equation}
  \Ftilde\left( k \right) = \frac{1}{\Nspatial} \,\sum\limits_{l=1}^{\Nspatial/2-1}\,\left( -1 \right)^{l+1}\,\frac{
    \sin\left( \frac{k \Nspatial}{2} \right)\,\sin\left( q_l \right)
  }{
    \cos\left( k \right) - \cos\left( q_l \right)
  }\,\Ftilde_l\,,
  \label{eq:Ftilde_k}
\end{equation}
where we introduce the shorthand $\Nspatial=L/a$ and the momentum $q_l$ is given by $q_l = 2\pi l / \Nspatial$. 
$\Ftilde_l = \Ftilde\left( q_l \right)$ denotes the $l$th Fourier mode and we assume $\Ftilde_l \approx - \Ftilde_{-l}$, which is true up to lattice artifacts and finite gauge statistics.

With the assumed form $\Ftilde(k) = k\,\ftilde(k^2)$ we immediately define the momentum-dependent form factor $\ftilde(k^2)$ from Eq.~(\ref{eq:Ftilde_k}) by dividing out the momentum $k$. For later reference, we divide by the lattice momentum $2\,\sin\left( k/2 \right)$ instead of $k$. These two momenta differ only by lattice artifacts and in particular lead to equivalent results at $k = 0$. We then get
\begin{equation}
  \ftilde\left( k^2 \right) = \frac{1}{\Nspatial} \,\sum\limits_{l=1}^{\Nspatial/2-1}\,\left( -1 \right)^{l+1}\,
%%%  
  \frac{
    \sin\left( \frac{k \Nspatial}{2} \right)
  }{
    \sin\left( \frac{k}{2} \right)
  }\,
%%%
  \cdot\frac{
    \sin\left( \frac{2\pi}{\Nspatial}l \right)
  }{
    \cos\left( k \right) - \cos\left( \frac{2\pi}{\Nspatial}l \right)
  }\,
%%%
  \Ftilde_l\,.
  \label{eq:ftilde_k}
\end{equation}
Eq.~(\ref{eq:ftilde_k}) can be evaluated at zero momentum, $k = 0$, and one gets a representation of the form factor at zero momentum transfer
\begin{equation}
  \ftilde\left( 0 \right) = \sum\limits_{l=1}^{\Nspatial/2-1}\,\left( -1 \right)^{l+1}\,
  \frac{\sin\left( \frac{2\pi}{\Nspatial}l \right)}{1 - \cos\left( \frac{2\pi}{\Nspatial}l \right)} \, \Ftilde_l
  = \sum\limits_{l=1}^{\Nspatial/2-1}\,\left( -1 \right)^{l+1} \cot\left( \frac{\pi}{\Nspatial} l \right)\, \Ftilde_l\,,
  \label{eq:ftilde_0}
\end{equation}
which is the result given in \cite{Wilcox:2002zt}. With the form $\Ftilde_l = q_l \,\ftilde_l$ and $\ftilde_l = \ftilde\left( q_l^2 \right)$ this leads to the representation of the form factor at zero momentum transfer
\begin{equation}
  \ftilde\left( 0 \right) = \sum\limits_{l=1}^{\Nspatial/2-1}\,\left( -1 \right)^{l+1}\,
%%%
  \left( \epow{iq_l/2} + \epow{-iq_l/2} \right)\, \ftilde_l\,,
  \label{eq:ftilde_0_2}
\end{equation}
where analogously to Eq.~(\ref{eq:ftilde_k}) instead of the continuum momentum $q_l$ we factored out the lattice momentum $2\sin(q_l/2)$.

\begin{figure}[<htpb]
  \centering
  \includegraphics[height=0.45\linewidth]{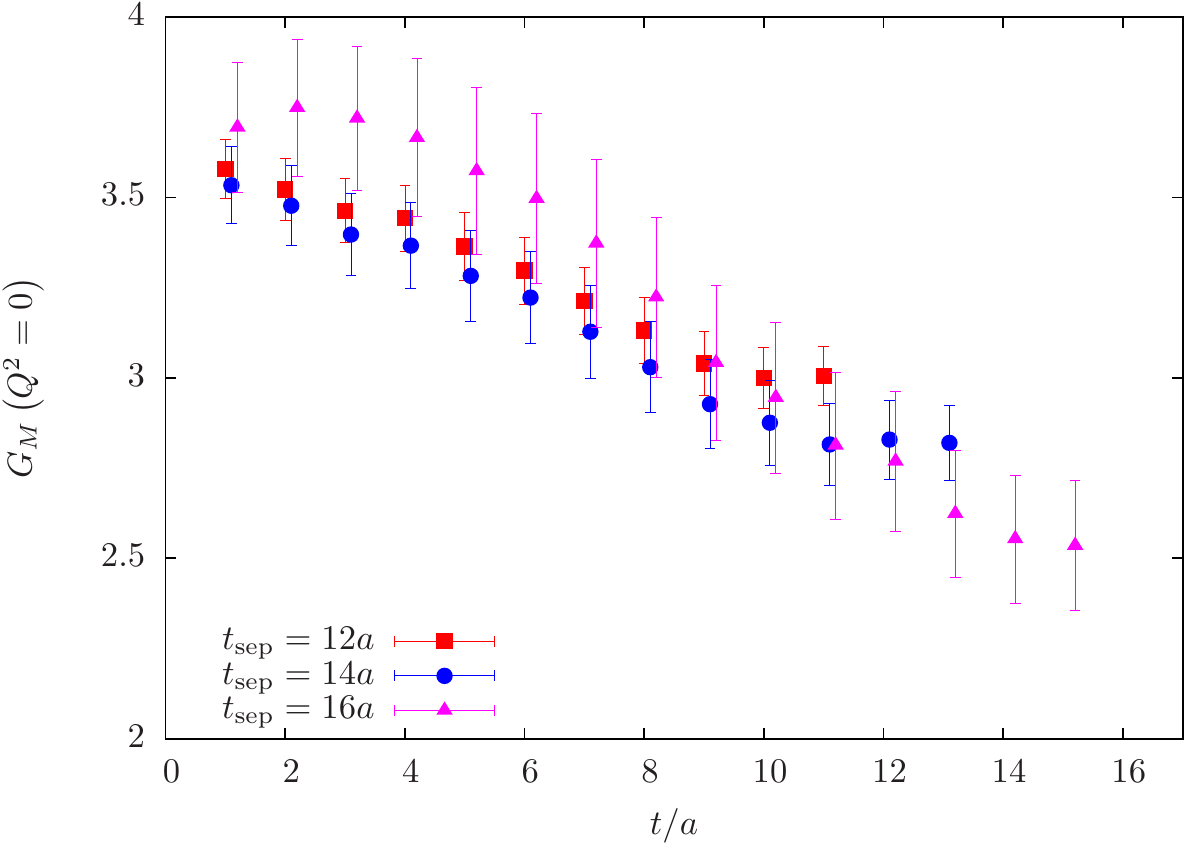}
  \caption{Residual dependence of the form factor at zero momentum transfer on the insertion time for different values of the source-sink separation.}
  \label{fig:GM_derivative_ratio}
\end{figure}

Equation~(\ref{eq:ftilde_0_2}) corresponds to a representation of the form factor at zero momentum as a superposition of Fourier-modes for nonzero momentum of the normalized 3-point function. We recall, that $\Ftilde_l\left( t \right) = \Ftilde(t,q_l)$ still depends on the current insertion time $t$. Inserting a complete set of states in Eq.~(\ref{eq:3pt_function_interpolator}) and assuming dominance of the nucleon state, the leading time dependence is found to be  $\Ftilde_l\left( t \right) \sim \Ftilde_l^{(0)} \,\epow{-t\left( E_N\left( q_l^2 \right) - m_N \right)}$. This renders $\ftilde_l$ a function of the insertion time as well
\begin{equation}
  \ftilde\left( 0 \right) = \sum\limits_{l=1}^{\Nspatial/2-1}\,\left( -1 \right)^{l+1}\,
%%%
  \left( \epow{iq_l/2} + \epow{-iq_l/2} \right)\, \ftilde_l^{(0)}\,\epow{-t\left( E_N\left( q_l^2 \right) - m_N \right)}\,.
  \label{eq:ftilde_0_3}
\end{equation}
This dependence on the insertion time is demonstrated in Fig.~\ref{fig:GM_derivative_ratio}, which is taken from Ref.~\cite{Alexandrou:2014exa}. Following Ref.~\cite{Wilcox:2002zt} and Eq.~(\ref{eq:ftilde_0_3}) we must not associate this time dependence to a single  Fourier mode. In particular taking the limit of large $t$ and retaining only the first mode $l = 1$, which has the slowest exponential decay $E_N \left( q_1^2 \right) - m_N = \sqrt{\left( 2\pi/L \right)^2 + m_N^2} - m_N$, leads to a wrong result.
The convergence of the sum in Eq.~(\ref{eq:ftilde_0_2}) with growing $\Nspatial$ is an essential point for the position space methods. A primary requirement is naturally, that the Fourier modes in Eq.~(\ref{eq:ftilde_0_2}) have an $l$-dependence, such that the sum is convergent in the limit $\Nspatial \to \infty$. For the example of a constant form factor $\ftilde_l = c = \mathrm{const}$ one has
\begin{equation}
  \ftilde\left( 0 \right) = c\,\left( 1 + \frac{\sin\left( \pi/\Nspatial \right)}{1 + \cos\left( \pi/\Nspatial \right)} \right) 
  = c\,\left( 1 - \tan\left( \frac{\pi}{2\Nspatial} \right) \right) \,,
  \label{eq:example_const}
\end{equation}
which converges to the target value with an error of $\mathcal{O}\left( 1 / \Nspatial \right)$ for large $\Nspatial$. As another example one may consider the case of a form factor $\ftilde_l = c\,\epow{-a|l|}$  with constants $a,c$, which leads to
\begin{equation}
  \ftilde\left( 0 \right) = c\,\left( 
  1 + \frac{\left( -1 \right)^{\Nspatial/2}\,\sin\left( \pi/\Nspatial \right)}{\epow{a\Nspatial/2}\left( \cos\left( \pi/\Nspatial \right) + \cosh\left( a \right) \right)}
  \right)\,,
  \label{eq:example_exp}
\end{equation}
and converges exponentially fast for positive $a$.

For the case at hand, we do not have the detailed momentum dependence for $\ftilde_l$, but we can employ e.g. the vector meson dominance (VMD) model
\begin{equation}
  \ftilde_l(t) \propto \frac{\epow{-t\left( E_N\left( q_l^2 \right) - m_N \right)}}{\left( 1 + Q^2_l / m_V^2 \right)^2} \,,
  \label{eq:ftilde_vmd}
\end{equation}
to investigate the convergence and the time dependence. Here $m_V$ is the mass of the exchanged vector meson and for the ensemble at hand we have $a m_V \approx 0.5$ in lattice units, which amounts to about $1\,\mathrm{GeV}$ in physical units at the given lattice spacing. $E_N{\left(q_l^2\right)}=\sqrt{q_l^2 + m_N^2}$ denotes the energy of the nucleon state on the mass shell for a given momentum $q_l$ and the four-momentum transfer $Q^2$ can be written as
\begin{equation}
  Q_l^2 = \left(p_f - p_i\right)^2 = 2m_N\left( E_N{\left( q_l^2 \right)} - m_N \right)
  \label{eq_four_momentum_transfer}
\end{equation}
for our setup with the momentum of the final state $p_f = \left( m_N, 0 \right)$ and momentum of the initial state $p_i = \left( E_N{\left( q_l^2 \right)},\, q_l\,\vec{e}_j \right)$.
\begin{figure}[htpb]
  \centering
  \subfigure[]{\includegraphics[width=0.48\textwidth]{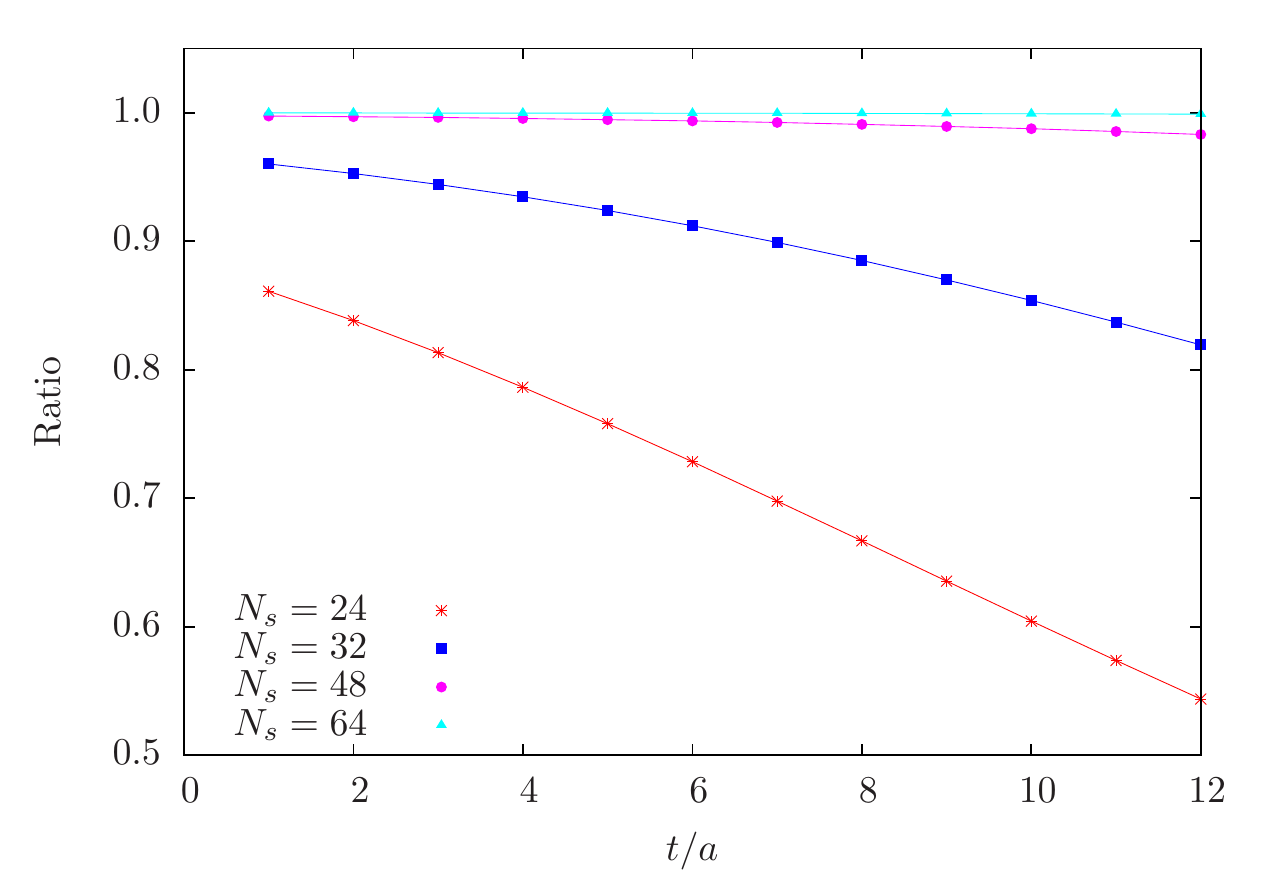}}\quad
  \subfigure[]{\includegraphics[width=0.48\textwidth]{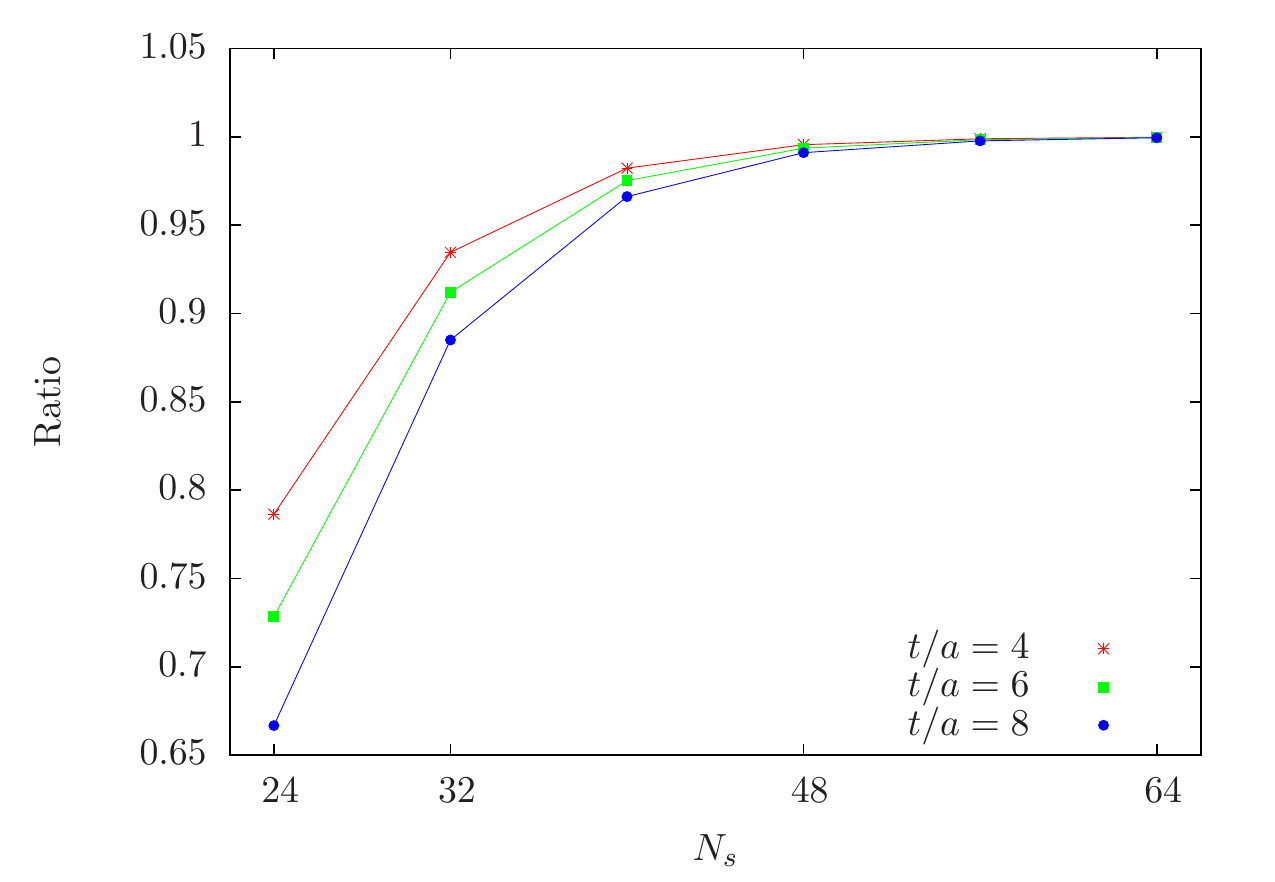}}
  \caption{Panel (a): residual dependence on the insertion time for the VMD model in Eq.~(\protect{\ref{eq:ftilde_vmd}}) with $a m_N = 0.5 = am_V$. Panel (b): dependence of the ratio evaluated at fixed current insertion times $t/a = 4,\,6,\,8$ on the size $\Nspatial$ for typical lattice sizes.}
  \label{fig:example_vmd_model}
\end{figure}
For the phenomenological VMD model in Eq.~(\ref{eq:ftilde_vmd}) the residual time dependence due to the time dependence of the individual Fourier modes is presented in the left plot of Fig.~\ref{fig:example_vmd_model}. The right plot shows the dependence on the lattice size for fixed current insertion times $t/a$. \par

The residual dependence on the insertion time $t$ discussed in the model calculation above poses an intricate problem, because at finite $L$ it obscures the additional $t$-dependence induced by the excited states. An analysis of the time dependence arising from excited states is required in order to isolate the ground-state contribution by establishing a region where the ratio is time-independent (plateau region). Any residual time dependence beyond the one arising from excited states will spoil the appearance of a plateau region and thus the extraction of the ground state matrix element. In the following we describe a modification of this method, which eliminates the residual time dependence by first isolating the ground state and then taking the derivative. \par

\subsection{Momentum elimination in the plateau-region}
While it is possible to deal with the issue of the correct order of finite volume and large Euclidean time limits for the continuum derivative method, the residual time dependence at any finite volume introduces a potentially large systematic error for any actual lattice calculation. Therefore, it seems beneficial to employ a different method that removes any explicit Euclidean time dependence right from the start. In the following we will discuss such a method that, while working at fixed, finite Euclidean times, automatically guarantees the right order of limits. We will see that this approach does not require any additional fit after the initial removal of the time-dependence, and thus for smaller systematic effects for currently available lattices.

The method is obtained by first performing a fit to the plateau in Eq.~(\ref{eq:ratio}) at each lattice momentum $\vec{q}$, hence removing the time dependence, before performing any manipulations in position space. This approach requires the usual assumption that the source-sink separation $t_\mathrm{sep}$ is large enough, such that excited state contaminations are sufficiently suppressed within the statistical error. This requirement is inherent in all lattice QCD computations of matrix elements and in practice is checked using different values of $t_\mathrm{sep}$. Thus, the first step is the same as the standard method for extracting data at nonzero momentum, which is then used e.g. for a dipole fit. However, while for the standard method we may simply average over all lattice momenta contributing to the same physical momentum before fitting the time dependence, this is not possible for this approach because Eq.~(\ref{eq:q_G_M}) contains a sum over momentum components $q_j$.

Let us first consider the case of on-axis momenta, i.e. $\vec{q}=(\pm q, 0,0)^T$ and all permutations thereof. Before we apply the fit to the plateau in Eq.~(\ref{eq:plateau}), we average over all momentum directions and contributing index combinations according to Eq.~(\ref{eq:q_G_M}) for a given value of the scalar momentum variable $q$. We denote the corresponding fitted ratios by $\Pi(q)$. 

In the next step we perform a Fourier transform to obtain a ratio $\Pi(y)$ in position space for which $\Pi(y) \approx -\Pi(-y)$ holds up to statistical fluctuations. In any actual lattice simulation a cutoff $q_\mathrm{max}$ is required such that only momenta with $q<q_\mathrm{max}$ enter the Fourier transform. The reason for this is a decreasing signal-to-noise ratio of $\Pi_\mu\l(\vec{q}, \Gamma_\nu\r)$ for increasing lattice momenta, which leads to a very noisy final result if such large momenta are included in the Fourier transform. In practice, it turns out to be reasonable to choose $q_\mathrm{max}$ corresponding to the lowest possible $Q^2$ for which the original ratio $\Pi_\mu\l(Q^2, \Gamma_\nu\r)$ is zero within statistical errors. Typically, this leads to a value of $q_\mathrm{max}$ which is much smaller than the maximal allowed lattice momentum. With $n=y/a$ we have
\begin{equation}
 \Pi(y)=\l\{\begin{array}{ll}
   +\Pi(n), & n=0,...,\Nspatial/2 \\
   -\Pi(\Nspatial-n), & n=\Nspatial/2+1,...,\Nspatial-1
   \end{array}\r. \,,
\end{equation}
where
\begin{equation}
  \Pi(n) = \frac{1}{6}\, \sum\limits_{i,j,k=1}^{3}\,\epsilon_{ijk}\,\sum\limits_{ |q_j| \le q_\mathrm{max}}\,\Pi_i\left( (q_j,q_i=0, q_k=0), \Gamma_k\right)\,\exp(iq_j n) 
  \nonumber
\end{equation}
and averaging over positive and negative values of $y$ we obtain an exactly antisymmetric expression $\overline{\Pi}(n)$ as can be inferred from Eqs.~(\ref{eq:q_G_E}, \ref{eq:q_G_M}). Finally, $\overline{\Pi}(n)$ is transformed back in a way that allows us to introduce continuous momenta. For the electric Sachs form factor the corresponding expression is
\begin{align} 
 \Pi(k) &= \l[\exp(ikn)\overline{\Pi}(n)\r]_{n=0,\,\Nspatial/2} + \sum\limits_{n=1}^{\Nspatial/2-1}\exp(ikn)\overline{\Pi}(n) + \sum\limits_{n=\Nspatial-1}^{\Nspatial/2+1}\exp(ik(\Nspatial - n))\overline{\Pi}(n) \notag \\
        &=\l[\exp(ikn)\overline{\Pi}(n)\r]_{n=0,\,\Nspatial/2} + 2i \sum\limits_{n=1}^{\Nspatial/2-1} \overline{\Pi}(n) \sin\l(\frac{k}{2}\cdot (2n) \r) \,.
\end{align}
Redefining the momentum variable $\hat{k} \equiv 2\sin\bigl(\frac{k}{2}\bigr)$ one finds
\begin{equation}
 \Pi(\hat{k})-\Pi(0) = i \sum\limits_{n=1}^{\Nspatial/2-1} \hat{k}\, P_n \,\bigl(\hat{k}^2\bigr) \overline{\Pi}(n) \,,
 \label{eq:Pnsum}
\end{equation}
where
\begin{equation}
 P_n\bigl(\hat{k}^2\bigr) = P_n\left(\left(2\sin\left(\frac{k}{2}\right)\right)^2\right)  = \frac{\sin(nk)}{ \sin\bigl(\frac{k}{2}\bigr)} \,, \\
\end{equation}
can be related to Chebyshev polynomials of the second kind and is an analytic function of $\hat{k}^2$ in $(-\infty, +4)$. This property allows for an evaluation of $\Pi(\hat{k})$ at any intermediate value. In the expression given in Eq.~(\ref{eq:Pnsum}) we can divide by $\hat{k}$. Using the current insertion and spin projection leading to Eq.~(\ref{eq:q_G_M}) as well as including the appropriate kinematic factors for $\Pi(n)$ we obtain the desired expression for the nucleon magnetic moment without explicit momentum factors
\begin{equation}
 G_M(\hat{k}^2) = i\sum\limits_{n=1}^{\Nspatial/2-1} P_n(\hat{k}^2) \, \overline{\Pi}(n) \,.
 \label{eq:method2}
\end{equation}
Note that in the limit $\hat{k}\rightarrow0$ division by $\hat{k}$ is equivalent to applying a derivative with respect to the continuous momentum variable $\hat{k}$. A similar expression for the electric Sachs form factor is obtained in exactly the same way using the current insertion and spin projection that lead to Eq.~(\ref{eq:q_G_E}) instead of the one leading to Eq.~(\ref{eq:q_G_M}). \par

Using again the VMD model albeit now with the dependence on the current insertion time $t$ eliminated, we can have a look at the dependence of the form factor $f\left( 0 \right)$ on
the lattice size $\Nspatial = L/a$ and the upper summation limit $l_\mathrm{max} = L/\left( 2\pi \right)\,q_\mathrm{max}$ of the Fourier modes in momentum space.
\begin{figure}[htpb]
  \centering
  \includegraphics[width=0.7\textwidth]{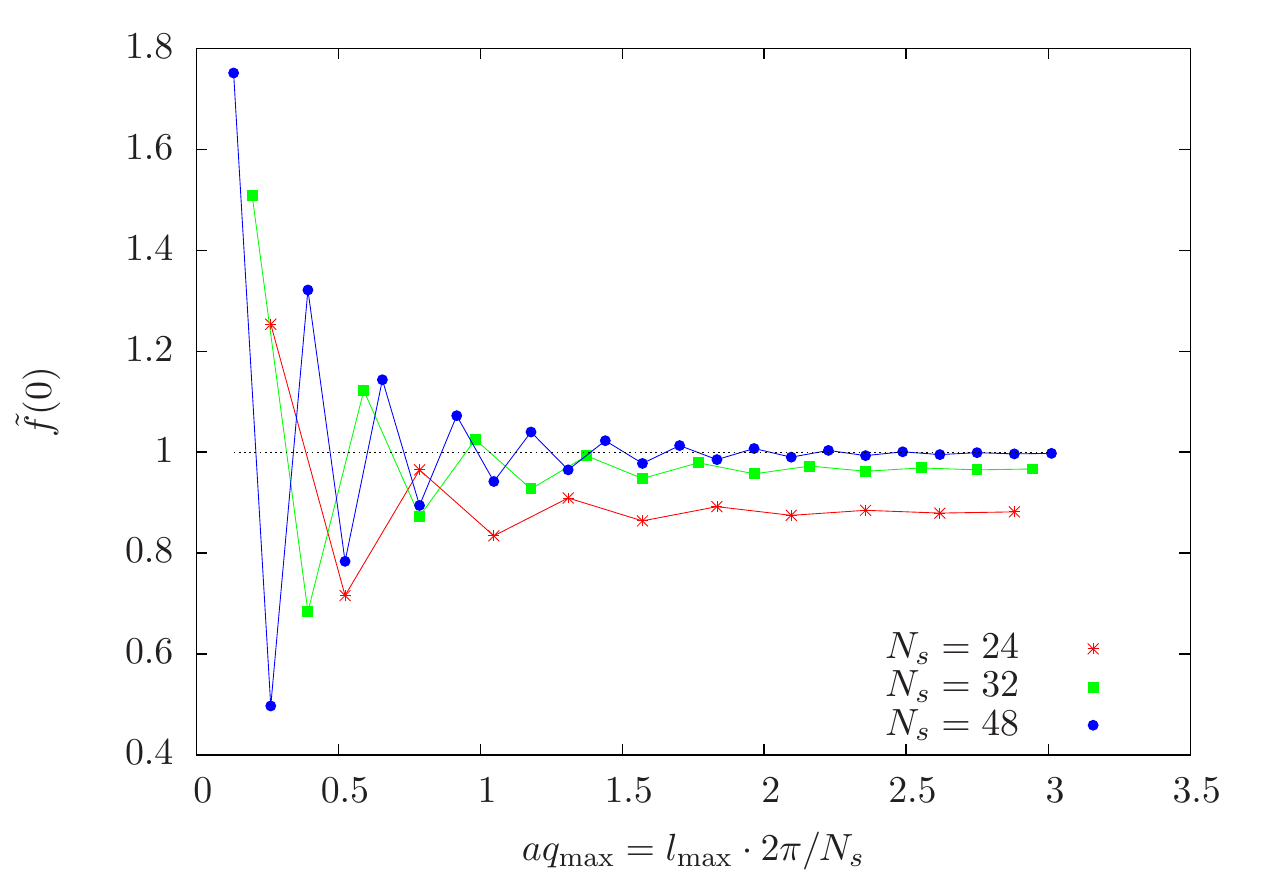}
  \caption{Dependence of the form factor $f\left( 0 \right)$ at zero four-momentum transfer on the cutoff $l_\mathrm{max} = L/\left( 2\pi \right)\,q_\mathrm{max}$ in momentum space for the VMD model.}
  \label{fig:lmax_dependence_vmd_m2}
\end{figure}
This behavior is depicted in Fig.~\ref{fig:lmax_dependence_vmd_m2} with the same model parameters as in subsection \ref{subsec:direct_application_of_the_momentum_derivative} above. For the VMD model we find the characteristic alternating convergence and an underestimation of the target value for too small $\Nspatial$. \par

We remark that it is straightforward to extend this method to arbitrary sets of off-axis momentum classes 
\begin{equation}
 M(q, q_\mathrm{off}^2) = \l\{\vec{q} \ | \ \vec{q}=\{\pm q, q_1, q_2\} \,, \ q_1^2+q_2^2=q_\mathrm{off}^2\r\} \,,
\end{equation}
where $\{\pm q, q_1, q_2\}$ denotes all permutations of $\pm q$, $q_1$ and $q_2$, with $ q=2\pi n/L$ for $n=0,...,\Nspatial/2$. However, to combine the results for $G_M(Q^2)$ for different $q_\mathrm{off}^2$--classes as a function of the now continuous Euclidean momenta $Q^2=Q^2(\hat{k}, q_\mathrm{off}^2)$ we need to consider an analytic continuation for classes with $q_\mathrm{off}^2>0$ to reach zero momentum, i.e. $Q^2=0$. This amounts to consistently replacing $k\rightarrow i\kappa$ and $\hat{k} \rightarrow i\hat{\kappa} = -2\sinh\bigl(\frac{\kappa}{2}\bigr)$ in the derivation outlined above. Note that in this case one has
\begin{equation}
 P_n\bigl(\hat{\kappa}^2\bigr) = \frac{\sinh(n\kappa)}{\sinh\bigl(\frac{\kappa}{2}\bigr)}\,.
\end{equation}
In principle, it is possible to combine results from momenta sets $M(q, q_\mathrm{off}^2)$ at different values of $q_\mathrm{off}^2$, e.g. in an error-weighted average. However, at finite values of the lattice spacing and finite lattice volumes, these momentum classes can be affected by different cutoff effects and may as well pick up different excited state contaminations, even at the same value of $t_\mathrm{sep}$. Furthermore, the achievable statistical error at zero momentum strongly depends on the value of $q_\mathrm{off}^2$. This is because for increasing values of $q_\mathrm{off}^2$, the lowest contributing momentum grows large, hence the region of extrapolation between $Q^2=0$ and the first contributing momentum becomes larger and less constrained by data. Besides, for larger values of $q_\mathrm{off}^2$, there are less momenta that can be used before the signal is lost in noise and the resulting extrapolation may become unstable and hence unreliable. Therefore, only a few classes with small values of $q_\mathrm{off}^2$ are expected to give any relevant contribution at $Q^2=0$ and one needs to carefully check if extrapolations from larger values of $q_\mathrm{off}^2$ can be used. In general, one expects the set of on-axis momenta $M(q,0)$ to give the best results with respect to the statistical error and possibly also systematic effects, as it contains the largest number of momenta, hence providing the most reliable extrapolation to zero momentum. 

Finally we remark that similar approaches using analytic continuation have been used in the context of calculating hadronic vacuum polarizations \cite{Feng:2013xsa,Feng:2013xqa}. We will refer to this approach as the \emph{ momentum elimination in the plateau-region method}. \par

\section{Lattice setup and analysis details}
For this exploratory study  we employ one gauge ensemble generated by the European Twisted Mass Collaboration (ETMC) with the Iwasaki gauge action \cite{Iwasaki:1985we} and $N_f=2+1+1$ dynamical quark flavors of Wilson twisted mass fermions \cite{Frezzotti:2000nk,Frezzotti:2003xj}. The simulations have been performed at maximal twist, such that automatic $\order{a}$ improvement is present \cite{Frezzotti:2003ni,Chiarappa:2006ae,Baron:2010bv}. The corresponding fermionic actions for a mass-degenerate, light quark doublet and a nondegenerate, heavy quark doublet read at maximal twist
\begin{align}
 S_F^l &= a^4\sum_x  \overline{\chi}^l(x)\bigl(D_W[U] + i \mu_l \gamma_5\tau^3  \bigr ) \chi^l(x) \,, \\
 S_F^h &= a^4\sum_x  \overline{\chi}^h(x)\bigl(D_W[U] + i \mu_\sigma \gamma_5\tau^1 + \tau^3\mu_\delta  \bigr ) \chi^h(x) \,, 
\end{align}
where $D_W$ denotes the massless Wilson-Dirac operator. The doublet fields $\chi^{l,h}(x)$, $\overline{\chi}^{l,h}(x)$ in the twisted basis are related to the physical fields $\psi^{l,h}(x)$, $\overline{\psi}^{l,h}(x)$ by
\begin{equation}
 \psi^{l,h}(x)=\frac{1}{\sqrt{2}}\left(1 + i \tau^3\gamma_5\right) \chi^{l,h}(x),\qquad {\rm and} \qquad \overline{\psi}^{l,h}(x)=\overline{\chi}^{l,h}(x) \frac{1}{\sqrt{2}}\left(1 + i \tau^3\gamma_5\right) \,.
\end{equation}

For the computation of two-point and three-point functions we employ the standard interpolating field for the nucleon,
\begin{equation}
 J_N^\alpha(x) = \epsilon^{abc} u^{a,\alpha} \left[ u(x)^{\top b} \mathcal{C}\gamma_5 d(x)^c\right] \,.
\end{equation}
where single-flavor quark fields are components of the doublet field $\psi^l(x)$ in the physical basis. In order to reduce the contribution of excited states and to increase the overlap with the nucleon ground state, we use Gaussian-smeared quark fields~\cite{Alexandrou:1992ti,Gusken:1989qx}. The corresponding smearing parameters $N_G=50$, $a_G=4$ have been optimized for the nucleon ground state~\cite{Alexandrou:2008tn}. Besides, we apply APE smearing to the gauge fields with parameters $N_\mathrm{APE}=20$ and $a_\mathrm{APE}=0.5$, respectively. \par

The ensemble used in this study has a lattice volume of $T/a\times(L/a)^3=64\times32^3$ and has been generated with a light quark mass corresponding to a charged pion mass of $M_\pi \approx 373 \mev$. The lattice spacing is given by $a\approx0.0823(10)\fm$ and has been determined from the nucleon mass~\cite{Alexandrou:2013joa}, which on the present lattice takes a physical value of $m_N = 1.220(5) \gev$. In the notation of Ref.~\cite{Baron:2011sf} this ensemble is denoted by $B55.32$ and we refer the reader to this reference for further information on the input parameters and simulation details. \par

All statistical errors are consistently computed from a jackknife analysis, although it turns out that effects from autocorrelation are negligible for the observables in this study. For the major part of this study we have used $4695$ gauge configurations and two source-sink separations $t_\mathrm{sep}/a = 12,14$ for the evaluation of the ratio $R_\mu(t_f,t,\vec{q},\Gamma_\nu)$ in Eq.~(\ref{eq:ratio}). In addition, we have considered $t_\mathrm{sep}/a=10$ and $t_\mathrm{sep}/a=16$ using $2429$ and $2263$ gauge configurations, respectively [cf. panel (a) of Fig.~\ref{fig:GM_iso}]. Adjacent gauge configurations are separated by two trajectories in the hybrid Monte-Carlo history. For the reduced statistics computations at $t_\mathrm{sep}/a=10,16$ the configurations are equally spaced over (almost) the entire available range of configuration. For the case of $t_\mathrm{sep}/a=16$ a possible doubling of statistics does not seem useful, as the expected statistical errors are still too large compared to $t_\mathrm{sep}/a=12,14$ for  our purposes. Therefore, we restrict our actual analysis for the magnetic moment to $t_\mathrm{sep}/a=10,12,14$ and use the results at $t_\mathrm{sep}/a=16$ for a consistency check. The fit ranges $[t_1, t_2]$ for the $t$-dependence of the plateau for the ratio in Eq.~(\ref{eq:plateau}) at different values of the source-sink separation are given in Table~\ref{tab:fit_ranges}. \par

\begin{table}[h]%done
 \centering
 \begin{tabular}{@{\extracolsep{\fill}}lcccc}
  \hline\hline
   $t_\mathrm{sep}/a$ & 10 & 12 & 14 & 16 \\
  \hline\hline
   $t_1/a$ & 2 & 3 & 4 & 5 \\
   $t_2/a$ & 8 & 9 & 10 & 11 \\
  \hline\hline
 \end{tabular}
 \caption{Fit ranges for the time-dependence of the plateau in Eq.~(\ref{eq:plateau}) for different values of the source-sink separation.}
 \label{tab:fit_ranges}
\end{table}

\section{Results}
\begin{figure}[t]
 \centering
 \subfigure[]{\includegraphics[height=.45\linewidth]{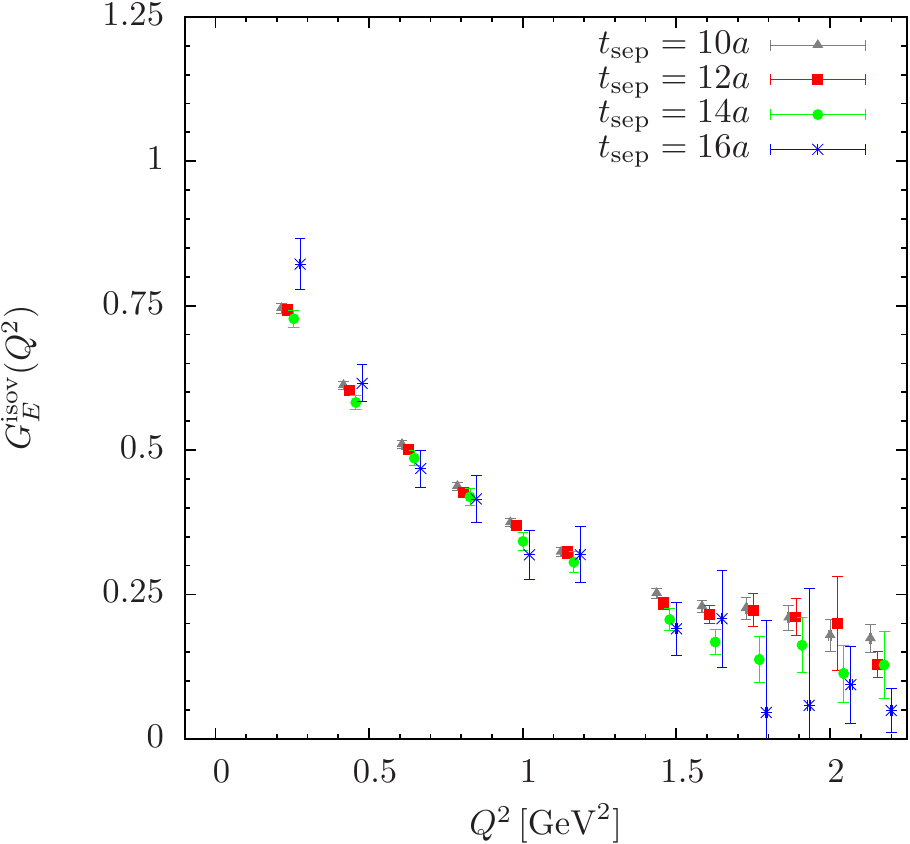}}\quad
 \subfigure[]{\includegraphics[,height=.45\linewidth]{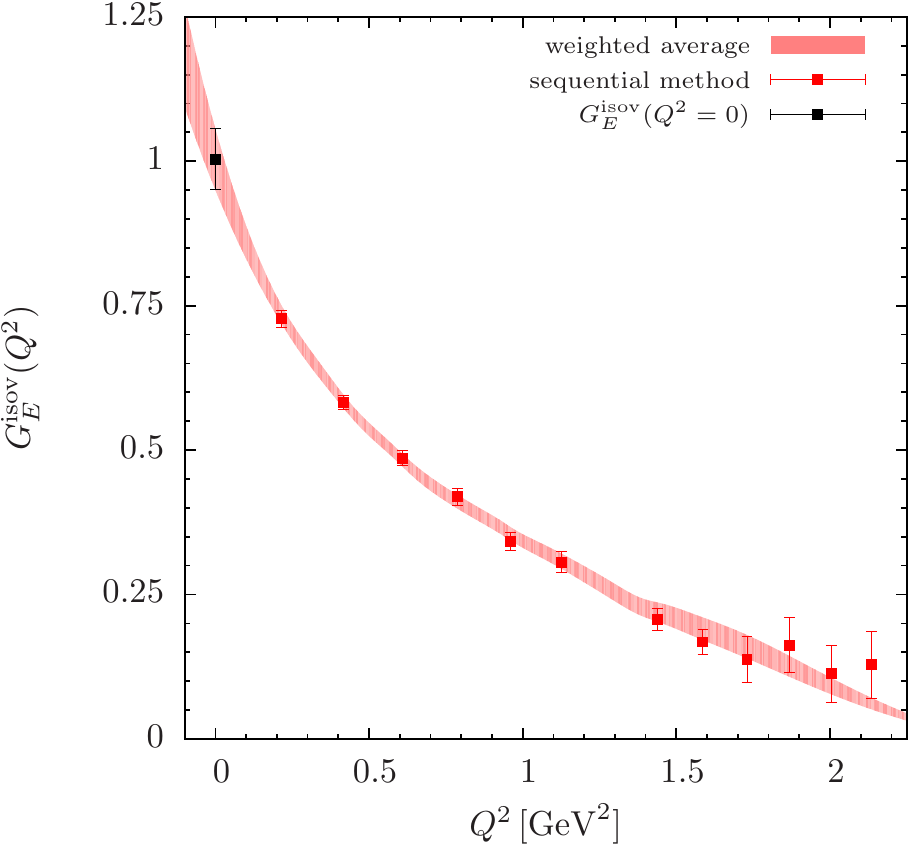}}
 \caption{Panel (a): Lattice data from the sequential method for the electric Sachs form factor $G_E^\iso(Q^2)$ on $B55.32$ at different source-sink separations $t_\mathrm{sep}/a$=10,12,14,16. Panel (b): Averaged results (red bands) for $G_E^\iso(Q^2)$ from momentum elimination in the plateau-region technique at $t_\mathrm{sep}/a=14$ taking the error-weighted average over the first five momentum sets $M(q, q_\mathrm{off}^2\leq5\cdot(2\pi/L)^2$. The resulting isovector electric form factor $G_E^\iso(0)$ is indicated by a black filled square. In addition, we show the corresponding lattice data points at discrete Euclidean momenta $Q^2$. Errors are statistical only.} 
 \label{fig:GE}
\end{figure}

As a first test of the momentum elimination  in the plateau-region method we consider the isovector electric Sachs form factor $G_E^\iso(Q^2)$ for which $G_E^\iso(0)=1$ holds exactly. To this end we start from the relation for $G_E^\iso(Q^2)$ in Eq.~(\ref{eq:q_G_E}), which exhibits a factor $q_j$ that prevents a direct computation at $Q^2=0$, unlike Eq.~(\ref{eq:G_E}), which is commonly used for a computation of the electric Sachs form factor. We collect in Table~\ref{tab:results_GE}  the results on $G_E^\iso(0)$ for all four source-sink separations. These results are obtained from the relation corresponding to Eq.~(\ref{eq:method2}) for the appropriate choice of spin projection an current insertion. The first data row contains results using only the on-axis momentum set $M(q,0)$. The  values for $t_\mathrm{sep}/a=10,12$ are larger than the actual value $G_E^\iso(0)=1$ while for $t_\mathrm{sep}/a=14$ the value agrees within the statistical error with unity. This effect is most likely due to residual excited state contaminations. Results for different source-sink time separations as a function of $Q^2$ are shown in the left panel of Fig.~\ref{fig:GE}.

\begin{table}[h]
\centering
 \begin{tabular}{@{\extracolsep{\fill}}lcccc}
  \hline\hline
   & $t_\mathrm{sep}/a=10$ & $t_\mathrm{sep}/a=12$ & $t_\mathrm{sep}/a=14$ & $t_\mathrm{sep}/a=16$ \\
  \hline\hline
   $G_E^\iso(0, q_\mathrm{off}^2=0)$ & 1.08(02) & 1.08(03) & 0.98(05) & 1.01(15) \\
   $G_E^\iso(0, q_\mathrm{off}^2\leq5\cdot(2\pi/L)^2)$  & 1.13(03) & 1.12(04) & 1.00(05) & 1.02(15) \\
  \hline\hline
 \end{tabular}
 \caption{Results for $G_E^\iso(0)$ at different values of $t_\mathrm{sep}/a$. First row contains results using on-axis momenta $M(q,0)$; second row those from error-weighted averaging over the first five momentum sets $M(q,q_\mathrm{off}^2\leq5\cdot(2\pi/L)^2))$. Errors are statistical only.}
 \label{tab:results_GE}
\end{table}

The second data row in Table~\ref{tab:results_GE} contains results for $G_E^\iso(0)$ from the error-weighted average over the first five sets of momentum classes $M(q, q_\mathrm{off}^2\leq5\cdot(2\pi/L)^2$. However, in practice it turns out that sets of momenta with $q^2_\mathrm{off}\geq2\cdot(2\pi/L)$ do not give any relevant contribution for the final value $Q^2=0$, because the corresponding statistical errors at $Q^2=0$ are  too large. Again, there is a similar trend visible from the results: For $t_\mathrm{sep}/a=10,12$ the results are too large and for $t_\mathrm{sep}/a=14$ we find agreement within errors. The extrapolation for $t_\mathrm{sep}/a=14$ is shown in the right panel of Fig.~\ref{fig:GE}. For the value at $t_\mathrm{sep}/a=16$ the statistical error is  too large for a meaningful statement, however the central value agrees with the result at $t_\mathrm{sep}/a=14$.

To include the off-axis momentum sets we require an extrapolation over a larger range in the $Q^2$ to reach zero momentum,
which  enhances  fluctuations and leads to larger statistical errors for the extrapolated value of $G_E^\iso(0)$.

\begin{figure}[t]
 \centering
 \subfigure[]{\includegraphics[height=.45\linewidth]{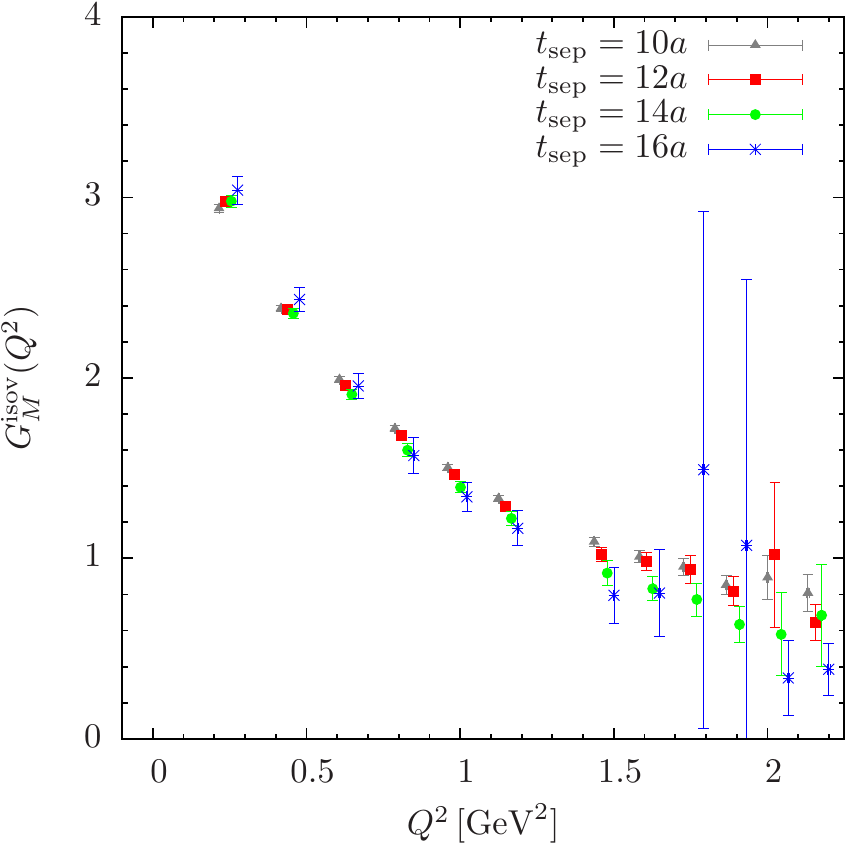}}\quad
 \subfigure[]{\includegraphics[height=.45\linewidth]{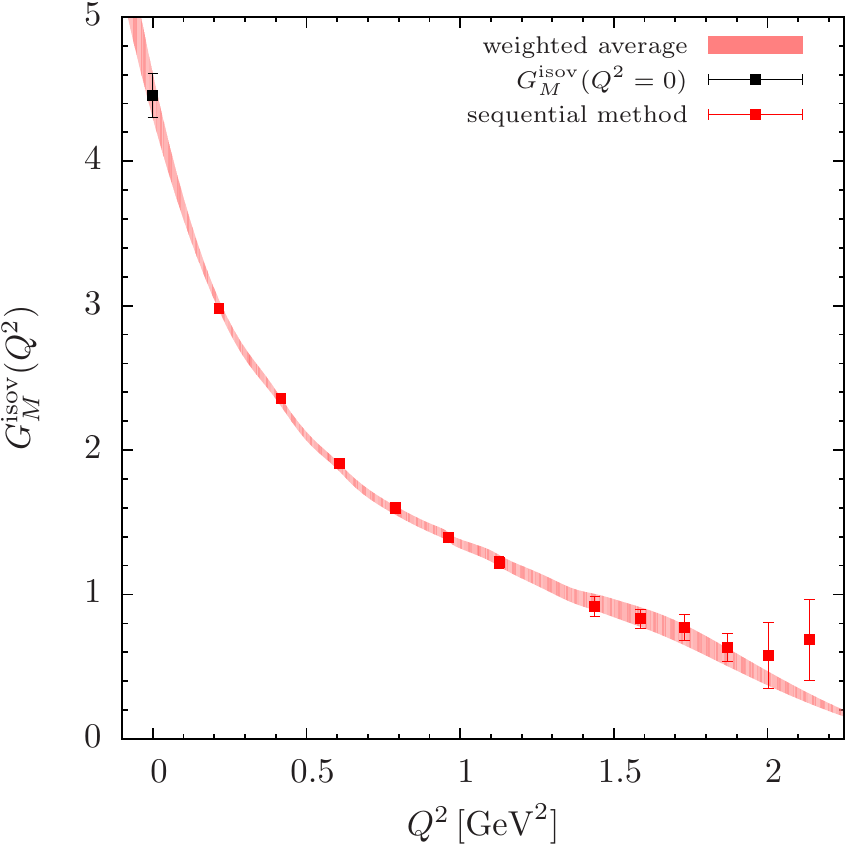}}
 \caption{Panel (a): Lattice data for $G_M^\iso(Q^2)$ on $B55.32$ for different source-sink separations $t_\mathrm{sep}/a$=10,12,14,16. Panel (b): Averaged results (red bands) for $G_M^\iso(Q^2)$ from the momentum elimination in plateau-region technique at $t_\mathrm{sep}/a=14$ taking the error-weighted average over the first five momentum sets $M(q, q_\mathrm{off}^2\leq5\cdot(2\pi/L)^2$. The resulting isovector magnetic form factor $G_M(0)$ is indicated by the black filled square. Errors are statistical only.}
 \label{fig:GM_iso}
\end{figure}

In the left panel of Fig.~\ref{fig:GM_iso} we show the lattice data for the isovector magnetic form factor of the nucleon. The data behave qualitatively similarly to those  for the electric form factor and again one finds residual excited states effect for smaller source-sink separations. The largest source-sink separation $t_\mathrm{sep}/a=16$ exhibits large statistical errors and it was not possible to use this data for a meaningful analysis, as it leads to large fluctuations and errors in the extrapolations. \par

\begin{figure}[t]
 \centering
 \subfigure[]{\includegraphics[height=.45\linewidth]{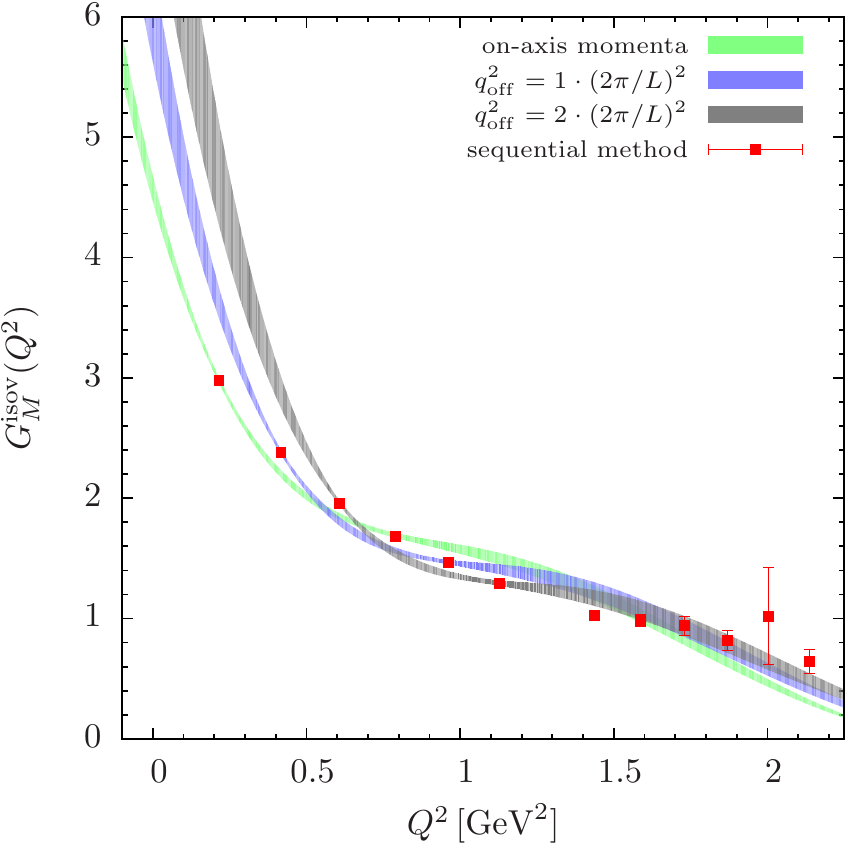}}\quad
 \subfigure[]{\includegraphics[height=.45\linewidth]{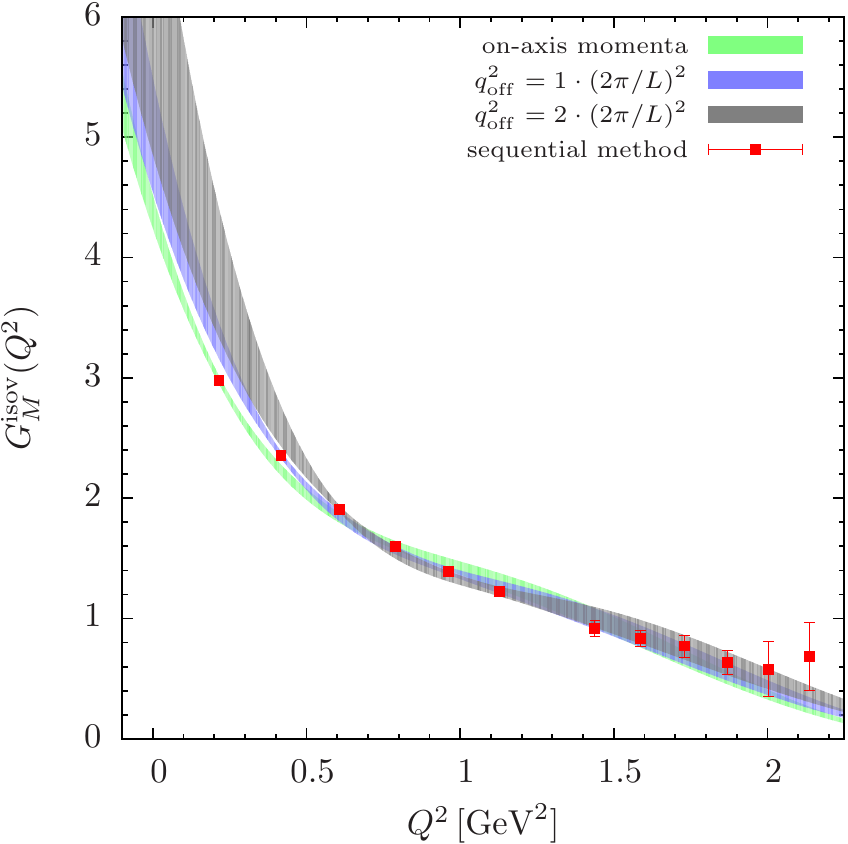}}
 \caption{Continuous results from the momentum elimination in plateau-region method for $G_M^\iso(Q^2)$ at (a) $t_\mathrm{sep}/a=12$ and (b) $t_\mathrm{sep}/a=14$ obtained from the first three momentum sets $M(q, q_\mathrm{off}^2\leq2\cdot(2\pi/L)^2$. In addition, we show the corresponding lattice data points at discrete Euclidean momenta $Q^2$.} 
 \label{fig:GM_iso_classes}
\end{figure}

Numerical results for the magnetic moment have been collected in Table~\ref{tab:results_GM}. Similar to the electric form factor we quote the results using only sets of on-axis momenta $M(q,0)$, as well as those from the error-weighted average over the first five sets $M(q,q_\mathrm{off}^2\leq5\cdot(2\pi/L)^2)$. The on-axis results are rather compatible within statistical errors although some effect due to excited states is expected for smaller values of $t_\mathrm{sep}/a$. 
In fact, for the averaged results such a trend is visible in the same way as for $G_E(0)$. This can also be seen from Fig.~\ref{fig:GM_iso_classes} where we show the first three sets of momenta $M(q,q_\mathrm{off}^2\leq2\cdot(2\pi/L)^2)$ separately: At $t_\mathrm{sep}/a=12$ (left panel) the extrapolations for different values $q_\mathrm{off}^2$ are clearly distinct, while at $t_\mathrm{sep}/a=14$ they are closer to each other albeit errors are larger as well.

For $t_\mathrm{sep}/a=14$ the results from on-axis momenta and averaging over $M(q,q_\mathrm{off}^2\leq5\cdot(2\pi/L)^2)$ agree within errors and we quote the averaged value at $t_\mathrm{sep}/a=14$ 
\begin{equation*}
 G_M^\iso(0) = 4.45(15)_\stat(07)_\sys \,, \notag
\end{equation*}
as our final result on this gauge ensemble. As a systematic error of our procedure we give the difference between on-axis and averaged results. The corresponding extrapolation at $t_\mathrm{sep}/a=14$ is also shown in the right panel of Fig.~\ref{fig:GM_iso}. Of course, a comparison to the experimental result $G_M^{\iso,\mathrm{exp}}(0) \approx 4.71$ \cite{Agashe:2014kda} must be considered with caution, as we did not perform a chiral or continuum extrapolation. Nevertheless, it is interesting that any value extracted from the momentum elimination in the plateau-region method turns out significantly larger than the result from fitting a dipole form to the data as given in Eq.~(\ref{eq:dipole_fit}). A previous study \cite{Alexandrou:2013joa} on the same gauge ensemble found a value of $G_M^\iso=3.93(12)$ at $t_\mathrm{sep}/a=12$ using a subset of 1200 gauge configurations. Here we have reanalyzed the full data set on the same configurations as used for the momentum elimination on the plateau method. The results are also given in Table~\ref{tab:results_GM} and the value for $G_M^\iso(0)$ is in good agreement with the one from 1200 configurations. \par

\begin{table}[h!]
\centering
\begin{tabular}{@{\extracolsep{\fill}}cccccccccc}
 \hline\hline
   & \multicolumn{3}{c}{$G_M^\iso(0)$} & \multicolumn{3}{c}{$G_M^\mathrm{p}(0)$} & \multicolumn{3}{c}{$G_M^\mathrm{n}(0)$} \\
  $t_\mathrm{sep}/a$ & $M(q,0)$ & average & dipole & $M(q,0)$ & average & dipole & $M(q,0)$ & average & dipole  \\
  \hline\hline
  10 & 4.47(07) & 4.63(08) & 3.76(03) & 2.76(04) & 2.85(05) & 2.32(02) & -1.71(03) & -1.77(03) & -1.44(01) \\
  12 & 4.57(11) & 4.70(12) & 3.89(03) & 2.81(06) & 2.89(07) & 2.40(02) & -1.76(04) & -1.81(05) & -1.50(01) \\
  14 & 4.38(14) & 4.45(15) & 4.01(04) & 2.69(08) & 2.73(09) & 2.47(03) & -1.69(06) & -1.72(06) & -1.54(02) \\
  \hline\hline
 \end{tabular}
 \caption{Collection of results for the magnetic moments $G_M^{\iso,\mathrm{p,n}}(0)$ at different values of the source-sink separation $t_\mathrm{sep}/a$. The results in subcolumns labeled $M(q,0)$ are obtained from on-axis momenta only, while those denoted by \emph{average}  contain results obtained from error-weighted averaging the separate results from the first five momentum sets $M(q,q_\mathrm{off}^2\leq5\cdot(2\pi/L)^2))$. In addition we have listed results from a dipole fit analysis using momenta up to $Q^2=1\mathrm{GeV}^2$. The results for the proton and neutron are obtained neglecting contributions from quark disconnected diagrams. Errors are statistical only.}
 \label{tab:results_GM}
\end{table}

Finally, we can also extract the magnetic form factors at $Q^2=0$ for the proton an the neutron. For the proton this is achieved by simply replacing the isovector by the electromagnetic current. In case of the neutron we make use of Eq.~(\ref{eq:isospin_current}) to obtain the relation
\begin{equation*}
 \bra{n} J^\mathrm{em}_\mu \ket{n} = \bra{p} J^\mathrm{em}_\mu \ket{p} - \bra{p} J^\iso_\mu \ket{p} \,,
\end{equation*}
which allows us to keep the proton interpolating field. Again, we employ the Noether current instead of the local current in the actual computation. As previously stated, a calculation for proton and neutron would involve quark disconnected diagrams, which we neglect for the purpose of the present study. It has been shown for this ensemble that the contribution of such diagrams to the magnetic moment is small~\cite{Abdel-Rehim:2013wlz}. This conclusion is also corroborated by a high-precision study using hierarchical probing~\cite{Green:2015wqa}. \par

\begin{figure}[t]
 \centering
 \subfigure[]{\includegraphics[height=.46\linewidth]{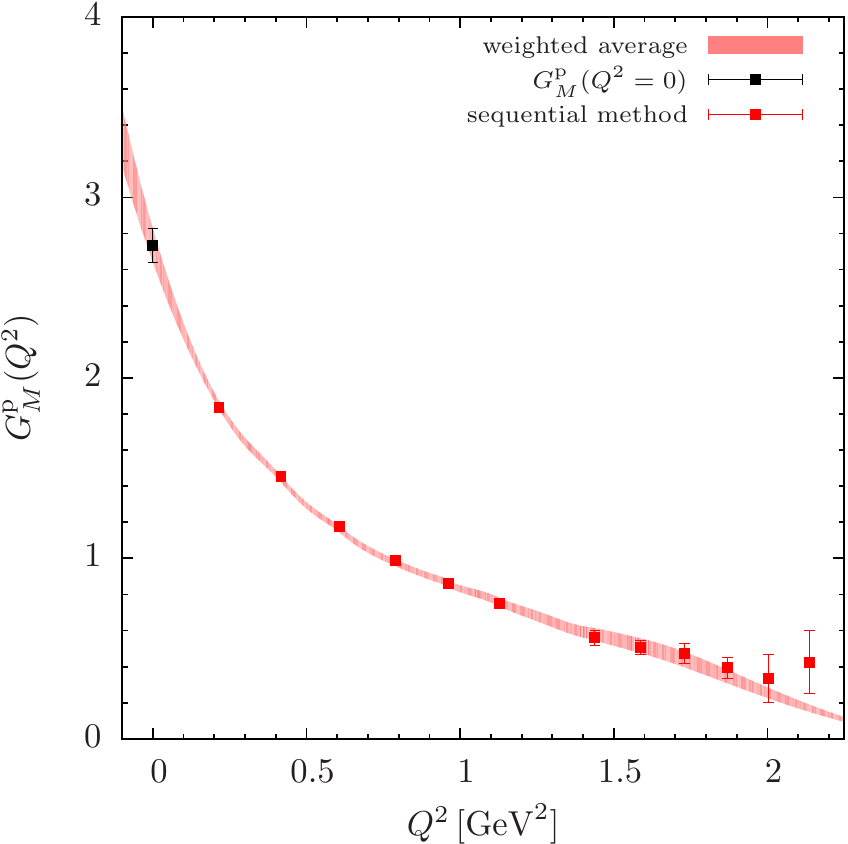}}\quad
 \subfigure[]{\includegraphics[,height=.46\linewidth]{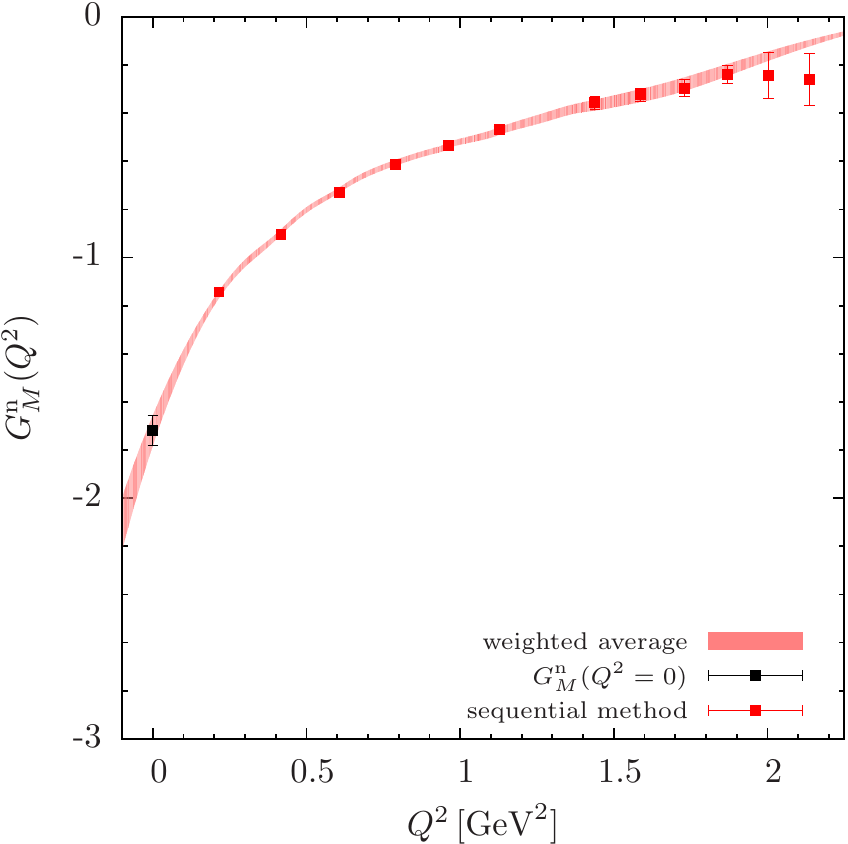}}
 \caption{Results for the magnetic moment of (a) the proton and (b) the neutron, obtained at $t_\mathrm{sep}/a=14$ averaging over momentum sets $M(q\leq5\cdot2\pi/L)$. }
 \label{fig:GM_pn_combined}
\end{figure}

The resulting values are also included in Table~\ref{tab:results_GM}. Similar to the isovector case we quote our final results from the averaged values at $t_\mathrm{sep}/a=14$
\begin{equation*}
 G_M^\mathrm{p}(0) = 2.73(9)_\stat(4)_\sys \quad \mathrm{and} \quad G_M^\mathrm{n}(0) = -1.72(6)_\stat(3)_\sys \,. \notag
\end{equation*}
For the proton we find surprisingly good agreement with the experimental value $G_M^\mathrm{p, exp}(0)\approx2.793$, while the experimental value for the neutron $G_M^\mathrm{n, exp}(0)\approx-1.913$ is more negative by three standard deviation as compared to  our result. In Fig.~\ref{fig:GM_pn_combined} we show the averaged extrapolation for $G_M^\mathrm{p}(0)$ and $G_M^\mathrm{p}(0)$ at $t_\mathrm{sep}/a=14$. Again, we find that results from a dipole fit give systematically smaller values, cf. Table~\ref{tab:results_GM}. \par

\section{Summary and Discussion}
The study of form factors within the lattice QCD formulation can provide nowadays a direct comparison to the experimental results since simulations with light quarks with physical values of their mass can be achieved. Beyond using simulations at the physical point, it is equally important to develop new methods to obtain more accurate results. In this study, we present an exploratory application of a new method to extract form factors directly at zero momentum transfer even if  they are multiplied in the matrix decomposition by kinematical factors of the momentum transfer. The basic idea is to eliminate the momentum factors using a method that is applicable for a finite lattice. This is accomplished by ensuring that excited states are sufficiently suppressed in the
ratio that yields the matrix element for large  Euclidean time evolution  and then take the {\it continuous derivative} of the Fourier transform of this ratio in position space.  This avoids the residual time dependence of a previous  method~\cite{Wilcox:2002zt} where the  derivative was applied to the three-point function written in terms of discrete momenta. The issue of this residual time-dependence was revisited  in Ref.~\cite{Alexandrou:2014exa} for the case of the magnetic form factor of the nucleon. In our current work we discuss in Sec.~3.1 in more detail  why this discrete derivative method fails and give the
correct order of taking the limits if it were to be applied. \par

In order to  demonstrate the applicability of the newly developed momentum elimination in the plateau-region method, we first apply it to extract the electric form factor at zero momentum transfer using Eq.~(\ref{eq:q_G_E}) where the same momentum prefactor appears as for the magnetic form factor in Eq.~(\ref{eq:q_G_M}). We show that our method correctly yields unity for the electric isovector form factor $G^\iso_E(0)$ and then apply it to extract the isovector magnetic form factor at zero momentum transfer. Using the fact that disconnected quark loop contributions are not larger than 1\% for this ensemble~\cite{Abdel-Rehim:2013wlz} we can extract the magnetic form factor for the proton and neutron. We find values that are closer to the experimental values as compared to what is extracted using the standard dipole fit. \par

We are currently extending our method for the direct extraction of the r.m.s. radii. Accomplishing this without using a fit ansatz will eliminate the model dependence in the momentum dependence of the form factors. Providing model-independent input from lattice QCD regarding the value of the electric radius of the proton can contribute to the understanding of the current discrepancy between muonic and electron scattering measurements.  We also plan to apply our method for simulation ensembles generated with the physical value of the light quark masses providing a direct comparison to the experimental measurements. \par

\section*{Acknowledgments}
We would like to thank all members of ETMC for the most enjoyable collaboration. Numerical calculations have used HPC resources from John von Neumann-Institute for Computing on the JUQUEEN and JUROPA systems at the research center in J\"ulich and on the Piz Daint machine at the Swiss National Supercomputing Center (CSCS) under project ID s540. The authors gratefully acknowledge the Gauss Centre for Supercomputing e.V. for funding the project pr74yo by providing computing time on the GCS Supercomputer SuperMUC at Leibniz Supercomputing Centre. Additional computational resources were provided by the Cy-Tera machine at The Cyprus Institute funded by the Cyprus Research Promotion Foundation (RPF), NEAY$\Pi$O$\Delta$OMH/$\Sigma$TPATH/0308/31.


\begin{thebibliography}{10}

\bibitem{Antognini:1900ns}
A.~Antognini {\em et~al.},
\newblock Science {\bf 339}, 417 (2013).

\bibitem{Mohr:2015ccw}
P.~J. Mohr, D.~B. Newell and B.~N. Taylor,
\newblock Rev. Mod. Phys. {\bf 88}, 035009 (2016),
  \href{http://arxiv.org/abs/1507.07956}{{\tt arXiv:1507.07956 [physics.atom-ph]}}.

\bibitem{Allison:2008xk}
I.~Allison {\em et~al.} (HPQCD Collaboration),
\newblock Phys. Rev. D{\bf~78}, 054513 (2008),
  \href{http://arxiv.org/abs/0805.2999}{{\tt arXiv:0805.2999 [hep-lat]}}.

\bibitem{Bernecker:2011gh}
D.~Bernecker and H.~B. Meyer,
\newblock Eur. Phys. J. A{\bf~47}, 148 (2011),
  \href{http://arxiv.org/abs/1107.4388}{{\tt arXiv:1107.4388 [hep-lat]}}.

\bibitem{Feng:2013xsa}
X.~Feng {\em et~al.},
\newblock Phys. Rev. D{\bf~88}, 034505 (2013),
  \href{http://arxiv.org/abs/1305.5878}{{\tt arXiv:1305.5878 [hep-lat]}}.

\bibitem{Francis:2013qna}
A.~Francis, B.~Jaeger, H.~B.~Meyer and H.~Wittig,
\newblock Phys. Rev. D{\bf~88}, 054502 (2013),
  \href{http://arxiv.org/abs/1306.2532}{{\tt arXiv:1306.2532 [hep-lat]}}.

\bibitem{Malak:2015sla}
Budapest-Marseille-Wuppertal Collaboration, R.~Malak {\em et al.},
\newblock  Proc. Sci. {\bf LATTICE2014} (2015) 161,
  \href{http://arxiv.org/abs/1502.02172}{{\tt arXiv:1502.02172 [hep-lat]}}.

\bibitem{Wilcox:2002zt}
W.~Wilcox,
\newblock Phys. Rev. D{\bf~66}, 017502 (2002),
  \href{http://arxiv.org/abs/hep-lat/0204024}{{\tt arXiv:hep-lat/0204024}}.

\bibitem{Alexandrou:2014exa}
C.~Alexandrou, M.~Constantinou, G.~Koutsou, K.~Ottnad and M.~Petschlies,
\newblock Proc. Sci. {\bf LATTICE2014} (2015) 144,
  \href{http://arxiv.org/abs/1410.8818}{{\tt arXiv:1410.8818 [hep-lat]}}.

\bibitem{Alexandrou:2015spa}
C.~Alexandrou {\em et~al.},
\newblock Phys. Rev. D{\bf~93}, 074503 (2016), 
  \href{http://arxiv.org/abs/1510.05823}{{\tt arXiv:1510.05823 [hep-lat]}}.

\bibitem{Alexandrou:2015ttm}
C.~Alexandrou {\em et~al.},
\newblock in {\em {Proceedings, 33rd International Symposium on Lattice Field Theory (Lattice 2015)}}, 2015,
  \href{http://arxiv.org/abs/1511.04942}{{\tt arXiv:1511.04942 [hep-lat]}}.

\bibitem{Dolgov:2002zm}
D.~Dolgov {\em et~al.} (LHPC collaboration, TXL Collaboration),
\newblock Phys. Rev. D{\bf~66}, 034506 (2002),
  \href{http://arxiv.org/abs/hep-lat/0201021}{{\tt arXiv:hep-lat/0201021}}.

\bibitem{Punjabi:2015bba}
V.~Punjabi, C.~F. Perdrisat, M.~K. Jones, E.~J. Brash and C.~E. Carlson,
\newblock Eur. Phys. J. A{\bf~51}, 79 (2015),
  \href{http://arxiv.org/abs/1503.01452}{{\tt arXiv:1503.01452 [nucl-ex]}}.

\bibitem{Feng:2013xqa}
X.~Feng {\em et~al.},
\newblock \href{http://arxiv.org/abs/1311.0652}{{\tt arXiv:1311.0652 [hep-lat]}}.

\bibitem{Iwasaki:1985we}
Y.~Iwasaki,
\newblock Nucl. Phys. {\bf B258}, 141 (1985).

\bibitem{Frezzotti:2000nk}
{\bf ALPHA} Collaboration, R.~Frezzotti, P.~A. Grassi, S.~Sint and P.~Weisz,
\newblock JHEP {\bf 08}, 058 (2001),
  \href{http://arxiv.org/abs/hep-lat/0101001}{{\tt hep-lat/0101001}}.

\bibitem{Frezzotti:2003xj}
R.~Frezzotti and G.~C. Rossi,
\newblock Nucl. Phys. Proc. Suppl. {\bf 128}, 193 (2004),
  \href{http://arxiv.org/abs/hep-lat/0311008}{{\tt hep-lat/0311008}}.

\bibitem{Frezzotti:2003ni}
R.~Frezzotti and G.~C. Rossi,
\newblock J. High Energy Phys. 08, (2004) 007,
  \href{http://arxiv.org/abs/hep-lat/0306014}{{\tt hep-lat/0306014}}.

\bibitem{Chiarappa:2006ae}
T.~Chiarappa {\em et~al.},
\newblock Eur. Phys. J. {\bf C50}, 373 (2007),
  \href{http://arxiv.org/abs/hep-lat/0606011}{{\tt arXiv:hep-lat/0606011 [hep-lat]}}.

\bibitem{Baron:2010bv}
R.~Baron {\em et~al.} (ETM Collaboration),
\newblock J. High Energy Phys. 06, (2010) 111, \href{http://arxiv.org/abs/1004.5284}{{\tt arXiv:1004.5284 [hep-lat]}}.

\bibitem{Alexandrou:1992ti}
C.~Alexandrou, S.~Gusken, F.~Jegerlehner, K.~Schilling and R.~Sommer,
\newblock Nucl. Phys. {\bf B414}, 815 (1994),
  \href{http://arxiv.org/abs/hep-lat/9211042}{{\tt arXiv:hep-lat/9211042 [hep-lat]}}.

\bibitem{Gusken:1989qx}
S.~Gusken,
\newblock Nucl. Phys. Proc. Suppl. {\bf 17}, 361 (1990).

\bibitem{Alexandrou:2008tn}
{\bf European Twisted Mass} Collaboration, C.~Alexandrou {\em et~al.},
\newblock Phys. Rev. D{\bf~78}, 014509 (2008),
  \href{http://arxiv.org/abs/0803.3190}{{\tt arXiv:0803.3190 [hep-lat]}}.

\bibitem{Alexandrou:2013joa}
C.~Alexandrou {\em et~al.},
\newblock Phys.Rev. D{\bf~88}, 014509 (2013),
  \href{http://arxiv.org/abs/1303.5979}{{\tt arXiv:1303.5979 [hep-lat]}}.

\bibitem{Baron:2011sf}
R.~Baron {\em et~al.},
\newblock Proc. Sci. {\bf LATTICE2010} (2010) 123,
  \href{http://arxiv.org/abs/1101.0518}{{\tt arXiv:1101.0518 [hep-lat]}}.

\bibitem{Agashe:2014kda}
K.~Olive {\em et~al.} (Particle Data Group Collaboration),
\newblock Chin. Phys. C{\bf~38}, 090001 (2014).

\bibitem{Abdel-Rehim:2013wlz}
A.~Abdel-Rehim {\em et~al.},
\newblock Phys. Rev. D{\bf~89}, 034501 (2014),
  \href{http://arxiv.org/abs/1310.6339}{{\tt arXiv:1310.6339 [hep-lat]}}.

\bibitem{Green:2015wqa}
J.~Green {\em et~al.},
\newblock Phys. Rev. D{\bf~92}, 031501 (2015),
  \href{http://arxiv.org/abs/1505.01803}{{\tt arXiv:1505.01803 [hep-lat]}}.


\end{thebibliography}
\end{document}